\documentclass{jfm}
\usepackage{graphicx}
\usepackage{epstopdf, epsfig}
\usepackage{amsmath}
\usepackage{here}

\title{Dynamics of unsteady premixed flames in meso scale channels and the effects of varying the wall heating conditions}

\author{
    Akhil Aravind\aff{1},
    Gautham Vadlamudi\aff{1}
    \and Saptarshi Basu\aff{1,2}
        \corresp{\email{sbasu@iisc.ac.in}}
    }

\affiliation
{\aff{1} Department of Mechanical Engineering, Indian Institute of Science, Bangalore, India.
\aff{2} Interdisciplinary Centre for Energy Research, Indian Institute of Science, Bangalore, India.}

\begin{document}

\maketitle

\begin{abstract}
Understanding the dynamics of flames at small scales opens up opportunities to enhance the performance of small-scale power generation devices, micro-reactors, fire safety devices, and numerous other systems that confine combustion to micro/meso scales. The current study delves into the dynamics of laminar premixed methane-air flames in mesoscale channels, subject to different wall heating conditions. Two external heaters, positioned at adjustable distances, are employed to create a bimodal wall heating profile on the combustor walls. The separation distance ($d$) between the heaters was varied, and the resulting flame dynamics were examined across a wide range of equivalence ratios ($\Phi$) and Reynolds numbers ($Re$). The observed dynamics were also compared against a baseline configuration that utilises a single heater to establish an axial unimodal wall temperature profile. Apart from the previously documented observations of unsteady flames with repetitive extinction and ignition (FREI) characteristics, this study identifies an additional unsteady propagating flame (PF) regime. While FREI appeared at stoichiometric and fuel-rich conditions, propagating flames were observed at the equivalence ratio of $0.8$. Unlike the FREI regime where the flame extinguishes after a characteristic travel distance, propagating flames continue to travel till they reach the upstream end of the combustor tube, where they extinguish upon encountering a meshed constriction. These flames are associated with a characteristic heat-release-rate oscillation that couples with the pressure fluctuations at frequencies close to the natural harmonic of the combustor tube. The observed dynamics of FREI and PF are discussed in detail with appropriate theoretical arguments to justify the observed trends. 
\end{abstract}


\section{Introduction}\label{sec:intro}
Rapid advances in micro-fabrication techniques have aided the miniaturization and integration of electro-mechanical systems on tiny chips. These micro-electro-mechanical systems (MEMS) find applications in areas ranging from biomedical devices to micro space/aerial vehicles. Currently, these devices rely on low-energy density batteries for their power requirements, which limits their portability and functionality. Micro/Meso combustors were initially conceived as potential alternatives to power these devices since the energy density of a typical liquid hydrocarbon is two orders higher in magnitude than standard alkaline/Li-ion batteries (\cite{fernandez-pello_micropower_2002,maruta_micro_2011,kaisare_review_2012}). However, micro/meso scale combustion has found its way into other domains in the form of micro-reactors (that find extensive applications in quantitative chemistry as reported by \cite{zimmermann_miniaturized_2002}) and fire safety systems. In this context, 'micro' describes combustors with a characteristic length scale smaller than the quenching diameter, while 'meso' combustors have a larger length scale that is of the order of the quenching diameter (\cite{ju_microscale_2011}).

When the size of the combustor decreases to micro or meso scales, its surface area to volume ratio increases significantly. Consequently, wall heat losses become significant, rendering the flame vulnerable to both thermal and radical quenching. To maintain flames at such scales, a commonly employed approach involves recirculating the product gases upstream of the reaction zone. This process preheats the reactants and modifies the temperature gradient at the wall to reduce heat losses, thus sustaining the flames in channels smaller than the classical quenching limit (\cite{lloyd_burner_1974}). In contrast to larger-scale combustors, where the combustor walls primarily serve as a sink for heat and radicals, the flame-wall interactions are more intricate at smaller scales. These interactions can, in fact, be such that the wall adds energy to the fluid, depending on the associated Biot and Fourier numbers. Such flame-wall coupling can thus introduce additional flame regimes that are otherwise not observed at large scales. Studies by \cite{evans_experimental_2011} report that the thermal properties of the walls, particularly thermal conductivity, exert a substantial impact on the temperature distribution along the wall and the overall behaviour of the flame. Furthermore, it has been observed that in small-scale combustors, elevated wall temperatures render the flame vulnerable to radical quenching (\cite{miesse_experimental_2005,kim_effects_2006,evans_experimental_2011}), potentially influencing the overall flame dynamics. The ensuing discussion delves into commonly observed flame behaviours at small scales.

\cite{maruta_characteristics_2005,maruta_characteristics_2004} conducted studies with a cylindrical quartz tube of inner diameter $2$ mm that acts as an optically accessible micro combustor for premixed methane-air flames. The effect of product gas re-circulation was imposed using two heating plates at the top and bottom of the tube. In addition to the stationary flames that establish themselves at a specific section within the channel, various unsteady flame behaviours were also documented. These included a flame exhibiting a series of ignition, propagation, extinction, and re-ignition events, referred to as FREI (Flames with Repetitive Extinction and Ignition), as well as a pulsating flame and a flame displaying traits of both pulsation and FREI. Similar observations were reported by \cite{fan_experimental_2009} in methane-air premixed flames inside rectangular quartz channels of different characteristic length scales, all of which were of the order of the quenching diameter. Likewise, \cite{richecoeur_experimental_2005} observed flame oscillations in curved mesoscale channels in fuel-rich conditions. \cite{ju_theoretical_2005} conducted both theoretical and experimental studies on the propagation and extinction of flames at mesoscales. They found that a flame within a mesoscale channel could propagate at velocities exceeding that of an adiabatic flame, contingent upon the thermal characteristics and heat capacity of the channel walls. Furthermore, it is to be noted that, alongside product gas re-circulation, stream-wise heat conduction along the wall from the flame was also found to affect the reactant temperature upstream of the flame front (\cite{kessler_ignition_2008}), the extent of which was again subject to the thermal properties of the wall. 

\cite{jackson_flames_2007} compared the problem of flame propagation/stabilization in narrow ducts to that of an edge flame and proposed a model that captured the transition between steady and unsteady flame solutions. The model only had thermal considerations and ignored hydrodynamic contributions by directly imposing a poiseuille velocity profile inside the duct. This treatment was further emphasized by \cite{bieri_effect_2011} and \cite{evans_experimental_2011}. Nonetheless, it was observed that the frequency of flame oscillations in unsteady flames aligned with the Strouhal number associated with the instability of the jet emerging from the micro/meso channels (St $\sim$ 0.4; \cite{richecoeur_experimental_2005}), emphasizing the role of hydrodynamics in dictating the quantitative flame dynamics. 

The Flame-wall interaction in unsteady flame regimes was extensively studied by \cite{evans_experimental_2011,evans_operational_2009}. Their work showed that a thin wall serves a dual purpose: acting as a heat sink following ignition, resulting in flame extinction due to rapid heat losses, as well as enabling re-ignition since the wall temperature rises before extinction due to the associated low thermal inertia. 

Consequently, this leads to high-frequency extinction and re-ignition events (FREI) and can result in acoustic emissions. Acoustic phenomena were typically observed at the lowest possible equivalence ratio that sustained oscillatory flames and occurred very close to the extinction phenomena. Similar observations of sound emission with unsteady flames were reported by various other research groups (\cite{evans_experimental_2011,evans_operational_2009, maruta_characteristics_2004, maruta_characteristics_2005, fan_experimental_2009, richecoeur_experimental_2005, MOHAN2020309}). \cite{nicoud_thermoacoustic_2005} attributed the sound generation to the contraction/expansion that happens during the extinction/re-ignition events associated with unsteady flames. The significance of such thermal effects was further emphasized by experimental results indicating that sound generation, or its absence, hinges primarily on the choice of wall material (\cite{evans_experimental_2011,evans_operational_2009}). The greater the thermal conductivity of the wall material, the more pronounced the sound emission becomes. \cite{d_p_experimental_2012} emphasized that this phenomenon is contingent on the specific apparatus used, and identified its origin as homogeneous explosions occurring at the tube entrance during the re-ignition process. Nevertheless, the exact cause of these acoustic phenomena associated with unsteady flames remains inconclusive.

Studies have also explored the micro/meso channel flame dynamics under different wall temperature profiles (all the investigated temperature profiles are monotonic along the combustor axis). \cite{kang_numerical_2017} numerically investigated the dynamics of methane-air flames in microchannels under different wall temperature conditions and concluded that the flame propagation speed was not only dependent on the chemical heat release rate but also on the flame-wall heat exchange rate. Consequently, these factors collectively led to quantitative changes in the characteristics of the flame regimes at different wall temperature profiles. A similar study by \cite{ratnakishore_dynamics_2017} demonstrated the movement of the ignition location to low wall temperature zones as the imposed axial wall temperature gradients were reduced. 

Building on the studies mentioned above, this work investigates the dynamics of flames in mesoscale channels subjected to a bimodal wall heating pattern. This scenario is of practical significance in mesoscale rectangular Swiss-roll burners, where hot product gases follow a rectangular outward spiral path while reactants travel along an inward rectangular spiral path to reach the reaction-stabilized centre of the combustor. The wall temperature is expected to decrease monotonically as we move radially outward from the reaction zone. However, as the flow decelerates at the corners of these rectangular spiral paths, regions with minor spikes in the wall temperature profile are anticipated.

The existing literature provides compelling evidence that such spikes in the wall temperature profile can significantly influence flame-wall interactions, leading to both quantitative and qualitative changes in flame dynamics. The primary objective of our research is to thoroughly investigate this phenomenon and enhance our understanding of the associated flame dynamics. As a first step, we explore the behaviour of unsteady premixed flames in channels featuring a bimodal wall heating profile. This bimodal profile represents a simplified one-dimensional depiction of the local temperature spike imposed over a monotonically decaying wall temperature profile along the combustor axis in the upstream direction.

\section{Experimental Setup}\label{sec:Expt_setup}

    \subsection{Mesoscale combustor facility} \label{sec:Test_facility}

\begin{figure}
        \centering
        \includegraphics [width=0.9\linewidth] {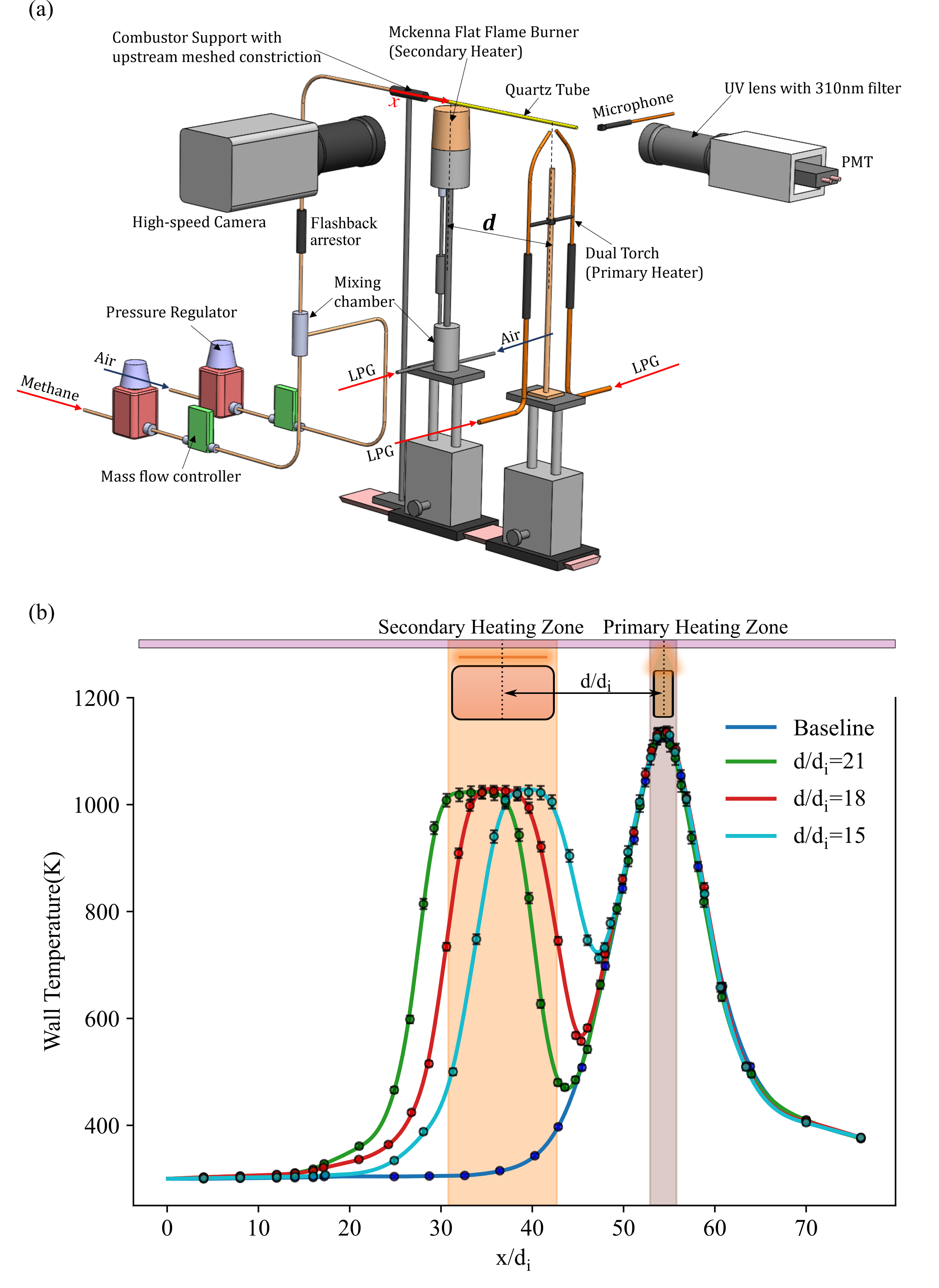}
        \caption{(a) Experimental setup (b) Inner wall temperature profiles at different separation distances ($d/d_{i}$). The regions highlighted in brown and yellow represent the primary and secondary heating zones, respectively, for $d/d_{i}=18$. (Color: Online)}
        \label{fig:Expt_Setup_Temp}
    \end{figure}

    A cylindrical quartz channel of $5$ mm inner diameter ($d_{i}$) and $7$ mm outer diameter ($d_{o}$) is used as an optically accessible mesoscale combustor (Length: $380$ mm). The channel is heated using two external heaters: a dual torch (primary heater) and a Mckenna flat flame burner (secondary heater), as depicted in Fig.\ref{fig:Expt_Setup_Temp}(a). The segment of the combustor tube directly above the primary and secondary heaters, which are referred to as the primary and secondary heating zones, respectively, is subject to a sharp rise and fall in wall temperature (along the combustor axis), peaking around $~1130K$ in the primary heating zone and around $~1025K$ in the secondary heating zone. The span over which the peak is maintained in the primary heating zone is relatively narrow compared to that observed in the secondary heating zone (Fig.\ref{fig:Expt_Setup_Temp}(b)), owing to the geometry of the heaters used. While the dual torch was confined to a narrow physical dimension, spanning $12$ mm along the combustor axis, the flame from the Mckenna burner had a wider span of approximately $60$ mm. In combination, the primary and secondary heaters impose a bimodal wall heating profile over the combustor walls. The distance between the centres of the heaters, characterised by $d$ (Fig.\ref{fig:Expt_Setup_Temp}(a)), is varied to obtain different bimodal wall heating profiles.

    Fig.\ref{fig:Expt_Setup_Temp}(b) depicts the spatial profile of the wall temperature measured along the inner walls of the combustor tube, using a $1$ mm K-type thermocouple. It is to be noted that since the combustor tube is heated from below, the inner wall temperature is not uniform across the combustor cross-section. The lower periphery, in closer proximity to the heaters, experiences a higher wall temperature compared to the upper periphery. This disparity was most pronounced in the primary heating zone, reaching approximately $2\%$ of the measured temperature. The temperature profiles depicted in Fig. \ref{fig:Expt_Setup_Temp}(b) represent inner wall temperature measured along the lower periphery of the combustor in the axial direction. The profiles are estimated from three sets of experimental measurements and correspond to quiescent conditions inside the combustor. 
    
    Methane and air regulated through two precise mass flow controllers (Bronkhorst Flexi-Flow Compact with the range of 0-1.6 SLPM for $CH_4$ and 0-2 SLPM for air) are directed into a mixing chamber, where the streams mix into each other to create a homogeneous mixture, which is then subsequently fed into the quartz combustor tube at its upstream end. The flow rates of LPG (liquefied petroleum gas) and air into the primary and secondary heaters (primary and secondary) were controlled using precise pressure regulators and mass flow controllers (Alicat Scientific MCR-500SLPM), respectively. In the discussions that follow, the $x$-axis is oriented along the combustor axis, extending along the downstream direction. The origin ($x=0$ mm) is set at the upstream end of the tube. The quartz tube connects to the upstream mixing chamber via a $1.5$ mm tubular constriction followed by a flashback arrestor. The $1.5$ mm tube houses a $100 \mu m$ wire mesh upstream of $x=0$ mm. 
    
    The current study was performed at three different equivalence ratios ($\Phi$): $0.8$, $1.0$, and $1.2$. The mixture velocity ($\bar{u}$) was varied between $0.1m/s$ and $0.3m/s$ in increments of $0.05m/s$, yielding upstream Reynolds numbers between $32$ and $96$ in increments of $16$. The experiments were conducted with four different wall heating conditions: A baseline case, wherein only the primary heater was used (no flat flame burner), and three other cases, wherein the distance between the centres of the dual-torch and the flat flame burner ($d$) was varied between $75$ mm to $105$ mm, in increments of $15$ mm, which corresponds to $d/d_{i}$ of $15$, $18$ and $21$. The axial temperature profile corresponding to these cases is plotted in Fig. \ref{fig:Expt_Setup_Temp}(b).

    \subsection{High-speed flame imaging and data acquisition from PMT and Microphone} \label{sec:Imaging}

    A Phantom Miro Lab 110 High-speed camera coupled with a $100$ mm Tokina macro-lens was used for high-speed flame imaging. The dynamics were captured at 4000 frames per second ($250 \mu s$ exposure time) with a spatial resolution of $200 \mu m$ per pixel (frame size of 1280 x 120 pixels). The data was used to track the spatial location of the flame. It is to be noted that the initial $65$ mm segment of the quartz tube was inaccessible for imaging due to the presence of a steel support that held the $380$ mm -long quartz combustor tube in the form of a cantilever (Fig. \ref{fig:Expt_Setup_Temp}(a)). The OH* chemiluminescence signal of the flame was captured using a Hamamatsu photomultiplier tube (H11526-110-NF). The PMT was positioned at a distance of $70$ mm from the downstream end of the combustor along the tube axis, such that the photocathode is exposed to the flame inside the combustor tube via a Nikon Rayfact (PF10445MF) UV lens and an OH* bandpass filter ($\sim 310$ nm); depicted in Fig.\ref{fig:Expt_Setup_Temp}(a). The pressure field fluctuation was recorded using a PCB microphone (PCB 130E20), which was placed at a radial distance of $80$ mm from the combustor axis at the downstream end of the combustor tube (Fig.\ref{fig:Expt_Setup_Temp}(a)). The data from the Photomultiplier tube (PMT) and the microphone was acquired using an NI-DAQ (PCI 6251) at 12000 Hz and was triggered alongside flame imaging via the high-speed camera.
    
    The flame images from the high-speed camera were processed in ImageJ, where they were subjected to thresholding using the Otsu thresholding technique, an integral feature of ImageJ. Otsu's thresholding algorithm calculates a single intensity threshold value ($I_f$) that separates all the pixels within an image into two categories: foreground and background. This threshold value, $(I_f)$, is determined by minimising the variance within each category or maximising the variance between the two. Pixels $(i, j)$ with intensities greater than or equal to $I_f$ are set to a binary value of 1, while those with intensities less than $I_f$ are assigned 0. The resulting binary area, comprising pixels with a value of 1, delineates the flame's boundary. Once the boundary is estimated, the location of the flame ($x_f$) is tracked by estimating the centroid of the region isolated by the flame boundary. The method has proven to be an effective technique for tracking the position of the flame and has been implemented previously by \cite{vadlamudi_insights_2021,pandey_self-tuning_2020,thirumalaikumaran_insight_2022} and \cite{Vadlamudi_Aravind_Basu_2023}. The position of the flame is tracked spatially with respect to time to obtain the flame propagation speed ($S_f$) in the reference frame of zero upstream mixture velocity, compensating for the relative velocity of the incoming mixture ($\bar{u}$) with respect to the propagating flame.
    
    \begin{gather*}
        S_f = \frac{dx_f}{dt} - \bar{u} \Rightarrow\ S_f = \frac{dx_f}{dt} + |\bar{u}| \tag{1}
    \end{gather*}
    
    It is to be noted that in the above relation, $\bar{u}$ is the velocity of the fuel-air mixture measured along the x-axis, while $S_f$ and $\frac {dx_f}{dt}$ are measured along the $-ve$ x-axis since the flame tends to propagate upstream with respect to the incoming flow. 
    
    The OH* chemiluminescence and microphone signals are processed in MATLAB after filtering it using the Savitzky-Golay filter (\cite{savitzky_smoothing_1964}), which is a low-pass filter based on the local least-square polynomial approximation that smooths the signal without distorting it (\cite{di_stazio_combustion_2016,di_stazio_oscillating_2016}). The data from the PMT and the pressure sensor were used to estimate the timescales associated with the unsteady meso-scale flame regimes. In the discussion presented in the subsequent sections, all the descriptors of unsteady flames (position, flame propagation speeds, OH chemiluminescence, frequency of repetition, time scales, etc.) are results averaged out over at least ten periodic repetition cycles from three different trials.

\section{Results and Discussions}\label{sec:results}

    \subsection{Global Observations}\label{Observation} \addvspace{10pt}

    Premixed methane-air mixture, at $300K$, enters the quartz combustor tube at $x=0$ mm and travels downstream, continuously gaining heat from the combustor walls and increasing its mean flow temperature ($T_m$). The mixture auto-ignites close to the primary heating zone where the inner wall temperatures are close to $\sim 1130K$. Upon auto-ignition, the mixture starts to propagate upstream, consuming the incoming reactants. This behaviour is consistently observed across the space of experimental conditions explored in the current work. However, this upstream propagating flame exhibits different dynamics contingent on the operating conditions of Reynolds numbers ($Re$), equivalence ratios ($\Phi$), and imposed wall heating profiles. Two global flame behaviours emerge, steady flames and unsteady flames. 

    \begin{figure}
        \centering
        \includegraphics [width=0.9\linewidth] {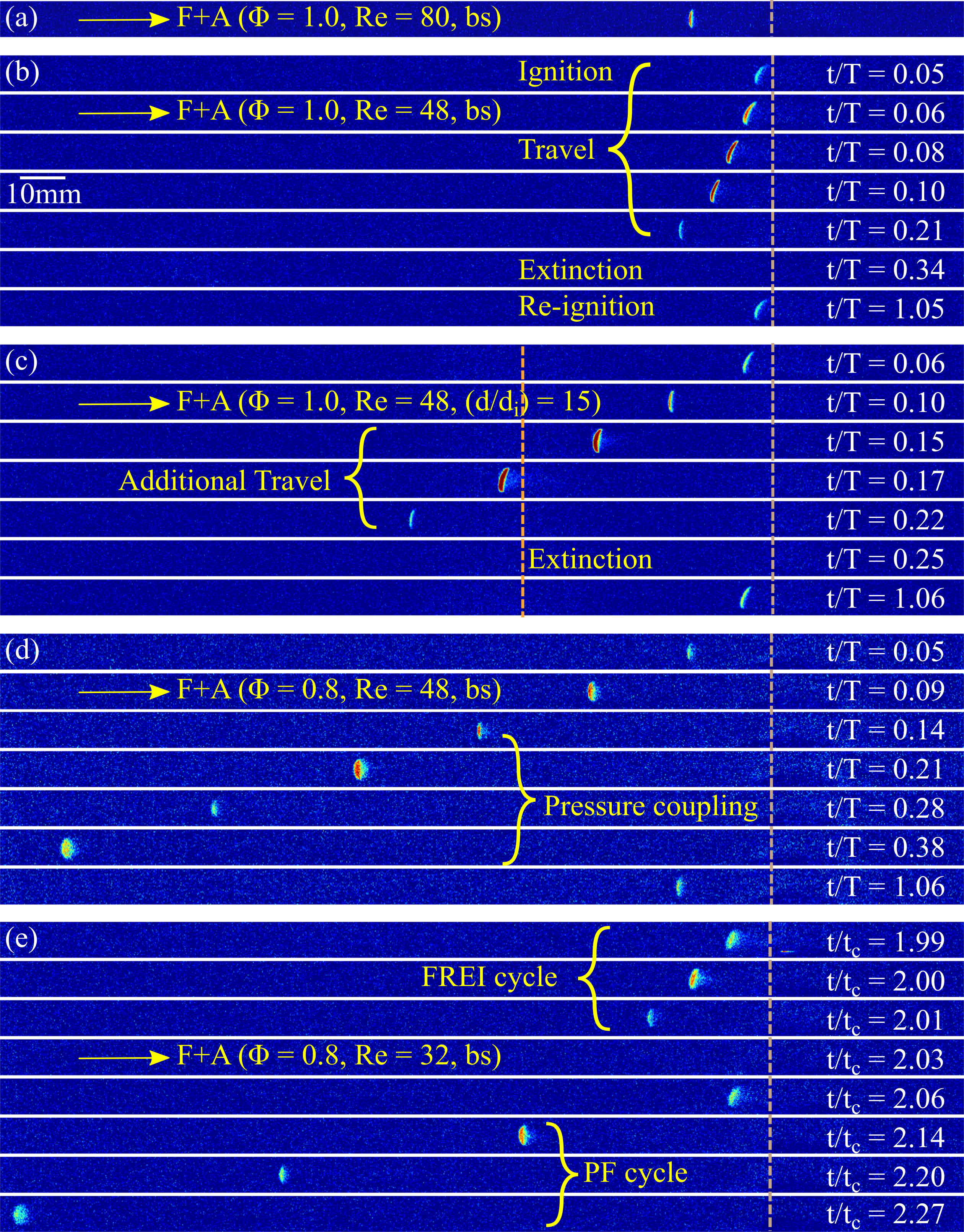}
        \caption{(a) Stationary Flames (SF). (b) Flames with repetitive extinction and ignition (FREI). (c) Diverging FREI (D-FREI). (d) Propagating Flames (PF). (e) Combined flame (CF). The brown and orange dashed vertical lines indicate the locations of the primary and secondary heaters, respectively. In the figure, $T$ is the characteristic time period of repetition of the unsteady flames, and $t_c$ is the convective time scale associated with the flow. Supplementary Movies: 1 to 4 illustrate the unsteady flame regimes depicted in Fig.\ref{fig:Obs_All_Flames}(b-e), respectively. Here, 'bs' denotes baseline conditions.}
        \label{fig:Obs_All_Flames}
    \end{figure}
    
    Stationary stable flames (steady flames) stabilise themselves at a characteristic upstream location (Fig. \ref{fig:Obs_All_Flames}(a)) post ignition. However, the unsteady flames demonstrate two distinct patterns; they either extinguish after traversing a characteristic distance (Fig. \ref{fig:Obs_All_Flames}(b)) or persist (continue propagating) until extinguished at the upstream meshed constriction of the combustor tube (Fig. \ref{fig:Obs_All_Flames}(d)) at $x=0$ mm. These unsteady flame regimes demonstrate periodic recurrence, reigniting after a characteristic time delay following extinction, as the fresh incoming mixture auto-ignites and repeat the flame cycle. Accordingly, three major flame regimes can be identified: Stationary Flames (SF), Flames with repetitive extinction and ignition (FREI), and Propagating flames (PF), respectively. In the subsequent sections, the dynamics of the latter-mentioned unsteady flame regimes (FREI and PF) are discussed, initially focusing on the trends observed in the baseline case (bs) and then comparing them with the changes observed due to the introduction of the secondary heater at different separation distances ($d$). Although stationary steady flames are identified in the present work, further experiments are necessary to establish conclusive trends in this regime since they (Stationary flames) are expected to sustain over a wide range of Reynolds numbers above $100$ (\cite{ju_microscale_2011}), which is beyond the scope of the present work.
   
    \begin{figure} 
        \centering
        \includegraphics [width=0.9\linewidth] {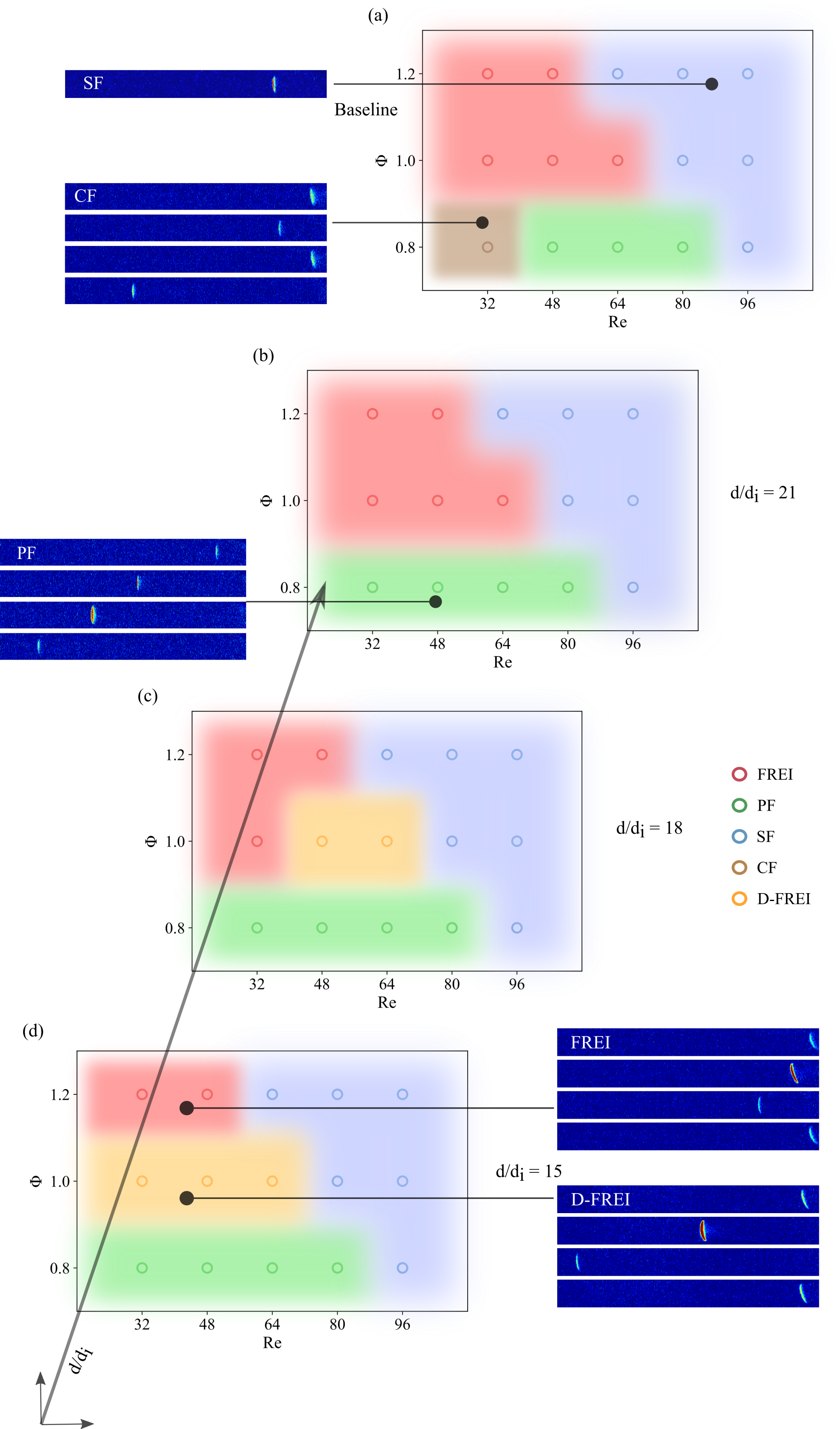}
        \caption{(a) Regime Map indicating the different flame regimes observed at the baseline case. (b-d) Regime Map corresponding to $d/d_{i}$ of $21$, $18$ and $15$, respectively.}
        \label{fig:Obs_Regime_Map}
    \end{figure}
    
    In the baseline configuration, the FREI regime appears at the equivalence ratio of $1.0$ and $1.2$, in the low Reynolds number regime, bounded by upper limits of $64$ and $48$, respectively (Fig.\ref{fig:Obs_Regime_Map}(a)). These flames, upon auto-ignition, propagate upstream and extinguish after a characteristic travel distance. This cycle is observed to repeat itself with a characteristic frequency ($\sim O (10) $), which increases with rising $Re$ or $\Phi$. Both the ignition and extinction locations are found to move downstream into regions of higher wall temperatures as the Reynolds number increases. Although the ignition locations are comparable between the equivalence ratios of $1.0$ and $1.2$, the flame tends to extinguish with a shorter flame travel at $\Phi=1.2$. 
    
    Introducing the secondary heater was found to qualitatively alter the OH* chemiluminescence and flame speed profiles of the FREI regime in a characteristic range of $Re$ and $\Phi$. At $\Phi=1.0$, the FREI dynamics changed significantly when the secondary heater was introduced at different separation distances ($d$). A $75$ mm separation between the heaters ($d/d_i$ = 15) was found to shift the flame’s extinction location upstream, increasing the flame’s travel distance while reducing the FREI repetition frequency. Additionally, an extra peak emerged in the OH* chemiluminescence and flame propagation speed ($S_f$) profiles. This regime that diverges from the baseline FREI behaviour will be referred to as the Diverging-FREI (D-FREI) in the sections that follow (Fig. \ref{fig:Obs_All_Flames}(c)). Similar trends were observed when the separation distance was increased to $90$ mm ($d/d_i$ = 18) for $Re \geq 48$, wherein the flame exhibited D-FREI behaviour. However, as the Reynolds number dropped below 48 (at $d/d_i$ = 18), the flame was found to retain the qualitative behaviour of the baseline FREI regime, exhibiting only minor quantitative variations in the FREI descriptors. Dynamics resembling the baseline FREI were also observed when the separation distance was further increased to $105$ mm ($d/d_i$ = 21) across the space of $Re$. However, unlike stoichiometric conditions, at the equivalence ratio of $1.2$, the dynamics were found to be weakly dependent on the separation distance ($d$), wherein the flame exhibited only minor quantitative variations (with respect to the baseline case) in the FREI descriptors for all values of $d/d_{i}$ (across the space of $Re$). These observations with relevant arguments are discussed in detail in sections \ref {FREI} and \ref{sec_heater}.
   
    In the baseline case, propagating flames (PF) were observed at the equivalence ratio of $0.8$ within the Reynolds number range of $48$ to $80$. PF, unlike FREI, continues travelling until it reaches the upstream end of the tube, wherein it is forced to extinguish by a meshed constriction (Fig.\ref{fig:Obs_All_Flames}(d)). Their ignition characteristics are similar to that observed in the FREI regime, wherein the ignition location moves downstream into regions of higher wall temperatures as $Re$ increases. However, unlike FREI, propagating flame develops instabilities during their propagation phase, that transform into violent back-and-forth motion of the flame front (Fig.\ref{fig:Obs_PF_Ins}(a)). These fluctuations are also found to be accompanied by a distinctive acoustic signal. A frequency domain analysis reveals that the fluctuations in the OH* chemiluminescence (heat release rate) and pressure signals are coupled during this phase and that the frequency of oscillation is close to the natural harmonic of the combustor tube. It is interesting to note that once the thermoacoustic coupling is established, the flame propagates upstream at a near-constant propagation velocity ($S_{f,ins}$), which tends to increase with an increase in the Reynolds number. In the presence of a secondary heater, the Reynolds number range over which the propagating flames were observed shifted between $32$ to $80$ for all values of $d/d_i$. The introduction was also found to elevate the peaks of the OH* chemiluminescence and the flame speed signals and quantitatively alter the flame descriptors. These trends are further discussed in section \ref{sec_heater}.
    
    A combined flame regime (CF) was identified at the equivalence ratio of $0.8$ and $Re=32$ (Fig.\ref{fig:Obs_All_Flames}(e)) in the baseline configuration. The flame exhibited characteristics of both FREI and PF, wherein a series of finite travel flame cycles (flame extinction after a characteristic travel distance, similar to FREI) was followed by a propagation flame cycle wherein the flame travelled up to the upstream end of the tube (similar to PF). The number of FREI cycles between consecutive PF cycles was stochastic. This flame regime is synonymous with transitional flame regimes that exist at the regime boundaries of different flame types, as reported \cite{ju_microscale_2011}. It is to be noted that this regime ceases to exist when the secondary heater is introduced and was only observed in the baseline configuration. 
    
    Figure \ref{fig:Obs_Regime_Map} presents a regime map that depicts the occurrence of the identified flame regimes as a function of equivalence ratio and Reynolds number at different wall heating conditions. In the sections that follow (\ref{FREI} and \ref{PF}), the above-mentioned characteristics of FREI and PF are discussed in detail with relevant scaling/mathematical arguments. However, to delve into such mathematical arguments, we need to first estimate the mean flow temperature ($T_m$) of the reactant mixture prior to ignition, which can act as a parameter to characterise and compare flame characteristics across the explored parametric space (Section \ref{Mean flow Temperature}).

    \subsection{Estimation of the mean flow temperature}\label{Mean flow Temperature} \addvspace{10pt}

    Mean flow temperature ($T_{m}$) is an estimate of the average temperature of the flow across the cross-sectional area (at a given axial distance, $x$) and can be used as a parameter to characterise the ignition-extinction location of the fuel-air mixture. The estimation of $T_m$ presented below is applicable only up to the section where the fluid packet auto-ignites or encounters the unsteady moving flame inside the tube. The following simplifying assumptions were used to evaluate $T_{m}$.
    
\begin{figure}
        \centering
        \includegraphics [width=0.7\linewidth] {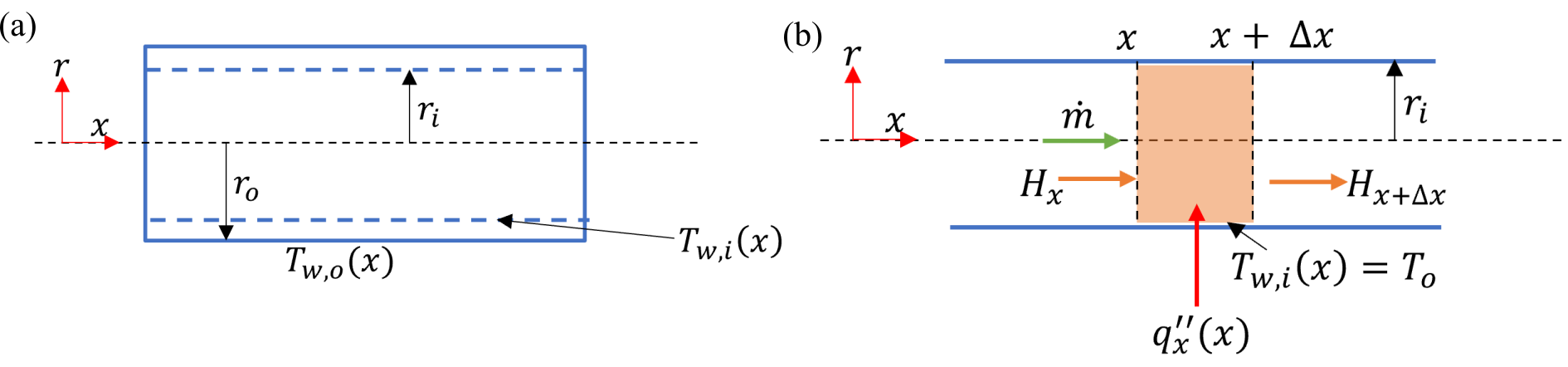}
        \caption{(a) Schematic depicting heat transfer across the tubular quartz combustor tube. (b) Schematic showing energy transfer across a control volume inside the quartz tube}
        \label{fig:Tm_Schematic}
    \end{figure}

    Experiments were conducted under steady-state external heating conditions. Due to the proximity of the external heaters to the combustor tube's lower periphery, a higher inner wall temperature is expected here compared to the upper periphery. However, since the disparities in the inner wall temperatures ($T_{w,i}(x,r)$) were measured to be $\leq 2\%$ of its measured value, we can approximate the wall temperature profile to have a radial uniformity. This implies that $T_{w,i}\left(x,r\right)=T_{w,i}\left(x\right)$ in Fig. \ref{fig:Tm_Schematic}(a). 
    
    As the unsteady regimes of FREI/PF propagate upstream following ignition, the flame interacts with the combustor walls. However, the wall-heating effects produced by these interactions can be considered negligible or short-lived, as the characteristics of the FREI/PF regimes (such as ignition location, flame speed, and OH* chemiluminescence signal) appear to remain unaffected by the wall-heating effects produced by these flames in their previous cycles. Therefore, we can disregard the wall-heating effects of the unsteady flames on the quartz tube and approximate the tube to remain in a steady state. Consequently, the energy balance equation for the quartz combustor tube simplifies to (Fig. \ref{fig:Tm_Schematic}(a)):

    \begin{gather*}
        \frac{\partial^2 T}{\partial r^2}+\frac{1}{r}\left(\frac{\partial T}{\partial r}\right)+\frac{\partial^2 T}{\partial x^2}=0 \tag{2}
    \end{gather*}
    
    It is to be noted that the above equation holds true for all Reynolds numbers and equivalence ratios of the premixture flow inside the quartz tube since the effect of the internal flow manifests in the form of boundary conditions along the inner walls of the tube. A simple scaling analysis can now be used to deduce the temperature drop across the inner and outer walls of the combustor tube.

    \begin{gather*}
        \frac{(\triangle T)_r}{(\triangle r)^2} \sim  \frac{(\triangle T)_x}{(\triangle x)^2} \tag{3}
    \end{gather*}

    The inner wall temperature profile, as measured using the thermocouple, reveals that the axial gradient of wall temperature is highest near the secondary heating zone wherein the temperature drops from $1025K$ to $300K$ ($(\triangle T)_x = 725K$) over an axial distance of $\sim 70$ mm ($\triangle x = 70$ mm). As per the above scaling law, this would imply that $(\triangle T)_r \sim 0.15K$, which would correspond to the highest temperature drop in the radial direction (taking $\triangle r \sim (r_{o}-r_{i})$, where $r_{o}$ and $r_{i}$ are outer and inner wall radius of the combustor tube respectively). We can thus approximate the inner and the outer wall temperatures to be comparable. Supplementary section S1 presents a finite difference formulation to estimate the outer wall temperature profile from the measured inner wall temperature profile. The plots in section S1 clearly depict that the fractional change in temperature between the inner and outer walls of the combustor tube is negligible. 

    Combining the above simplifications with the fact that the outer wall temperature is maintained constant by the external heaters (\cite{di_stazio_combustion_2016}), we can assume that the measured inner wall temperature profile remains temporally invariant and does not change with $Re$ and $\Phi$. Thus, the inner wall temperature profile in Fig.\ref{fig:Expt_Setup_Temp}(b), which was measured under quiescent conditions can be assumed to hold true across all experimental conditions explored in the current study. 
    
    We can now estimate the heat transferred from the combustor walls to the fluid moving inside the tube, assuming a temporally invariant inner wall temperature profile. For the Reynolds number range under consideration, the hydrodynamic entrance length is of the order of $O(10^{0})$ mm, which is negligible in comparison with the length of the tube ($380$ mm). We can thus assume the flow to be fully developed as it passes through the primary and secondary heating zones. Additionally, the peclet number associated with the flow is the order $O(10)$, and hence, the effect of axial conduction inside the flow can be neglected in comparison with axial advection effects (\cite{Bejan}). The energy balance equation for the steady reactant mixture stream moving downstream inside the combustor tube, thus, reduces to a balance between the radial conduction effects from the combustor walls and axial advection effects associated with the flow.

    \begin{gather*}
        \frac{u\left(r\right)}{\alpha}\left(\frac{\partial T}{\partial x}\right)=\frac{\partial^2T}{\partial r^2}+\frac{1}{r}\left(\frac{\partial T}{\partial r}\right) \tag{4}
    \end{gather*}
 
    A simple scaling analysis can be used to show that the Nusselt number will remain constant under these considerations (\cite{Bejan}). For an elemental control volume of length $\triangle x$ (where $\triangle x << L$; L: length of the combustor tube) in the flow domain (Fig. \ref{fig:Tm_Schematic}(b)), the wall temperature can be assumed to remain spatially constant across the elemental length of $\triangle x$. Invoking the assumption detailed earlier on the temporal invariance of the inner wall temperature, we can assume the inner wall temperature to remain locally constant across $\triangle x$ (spatially and temporally). This simplification helps in estimating the value of the Nusselt number to 3.66 (\cite{Bejan}).
    
    The energy balance energy can be further simplified to get,
    
    \begin{gather*}
        q^{''}_{w}(x)(2 \pi r_i \triangle x) = \dot{m} ( H(x + \triangle x) - H(x) ) \tag{5}
    \end{gather*}

    where $q^{''}_{w}(x)$ is the wall heat flux at the axial distance of $x$, $\dot{m}$ is the reactant mass flow rate, and $H(x)$ is the specific enthalpy of the mixture. Substituting for $q^{''}_{w}(x)$ as $h ( T_{o}(x) - T_{m}(x) )$, where $h$ is the co-efficient of heat transfer which is evaluated in terms of Nusselt number as, $h=((Nu) k))/(2 r_{i})$; and using $H(x) = C_{p}(T)T_{m}(x)$ in the above equation, we get,       

    \begin{gather*}
        T_m (x + \triangle x) = T_m (x) + \triangle x \Biggl( \frac{\alpha (Nu)}{\bar{u} r_{i}^2} ( T_o(x) - T_m(x) ) \Biggr) \tag{6}
    \end{gather*}
    
    In the above equation, $\alpha$ is the thermal diffusivity and $\bar{u}$ is the mean flow velocity of the mixture. Since the mean flow temperature at the inlet of the quartz tube ($x=0$ mm) is known, the above equation can be used to march spatially along the combustor axis to obtain the mean flow axial temperature profile. 

    \begin{figure} [h]
        \centering
        \includegraphics [width=0.9\linewidth] {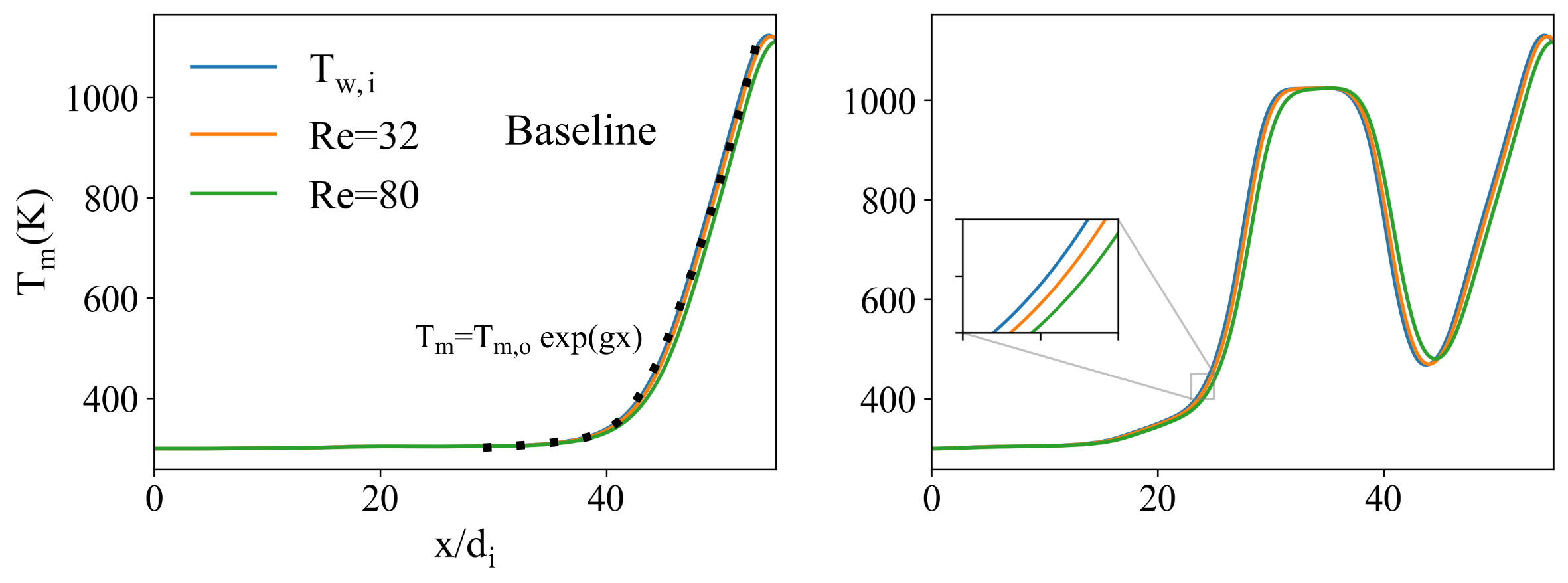}
        \caption{Panels (a,b) plots the mean flow temperature profiles in the baseline configuration and at $d/d_{i} = 21$, respectively. The plots correspond to the Reynolds numbers of $32$ and $80$ and are plotted alongside the inner wall temperature profile.}
        \label{fig:Tm_all}
    \end{figure}
    
    Fig. \ref{fig:Tm_all} plots the axial variation of the mean flow temperature for the baseline configuration and the case corresponding to $\frac{d}{d_i}=21$, at the Reynolds number of $32$ and $80$. In the range of $x/d_{i}$, where the wall temperature gradient is positive ($\frac{dT_{w,i}}{dx} > 0$), the plots show that the mean flow temperature drops as the Reynolds number increases, and the reverse is true when $\frac{dT_{w,i}}{dx} < 0$ (Fig.\ref{fig:Tm_all}(a,b)). In general, the mean flow temperature profile tends to shift downstream (with respect to $T_{w,i}$) with increasing Re, and this downstream shift becomes more pronounced at higher Reynolds numbers. The plots imply that in the regions where the temperature gradient is positive, the reactants have to travel longer distances along the combustor axis to reach a given mean flow temperature as $Re$ increases.
    
    Supplementary section S2 presents a numerical simulation accounting for the conjugate transfer between the combustor tube walls and the internal flow. The plots in the section demonstrate a strong agreement between the theoretical estimate of the mean flow temperature and the estimation of $T_m$ from the numerical simulations.

    \subsection{Baseline Configuration}\label{Baseline}
    This section explores the dynamics of the unsteady flames regimes across the space of $Re$ and $\Phi$ in the baseline configuration.
    
    \subsubsection{Flames with Repetitive Extinction and Ignition (FREI)}\label{FREI} \addvspace{10pt}
   
\begin{figure}
            \centering
            \includegraphics [width=0.9\linewidth] {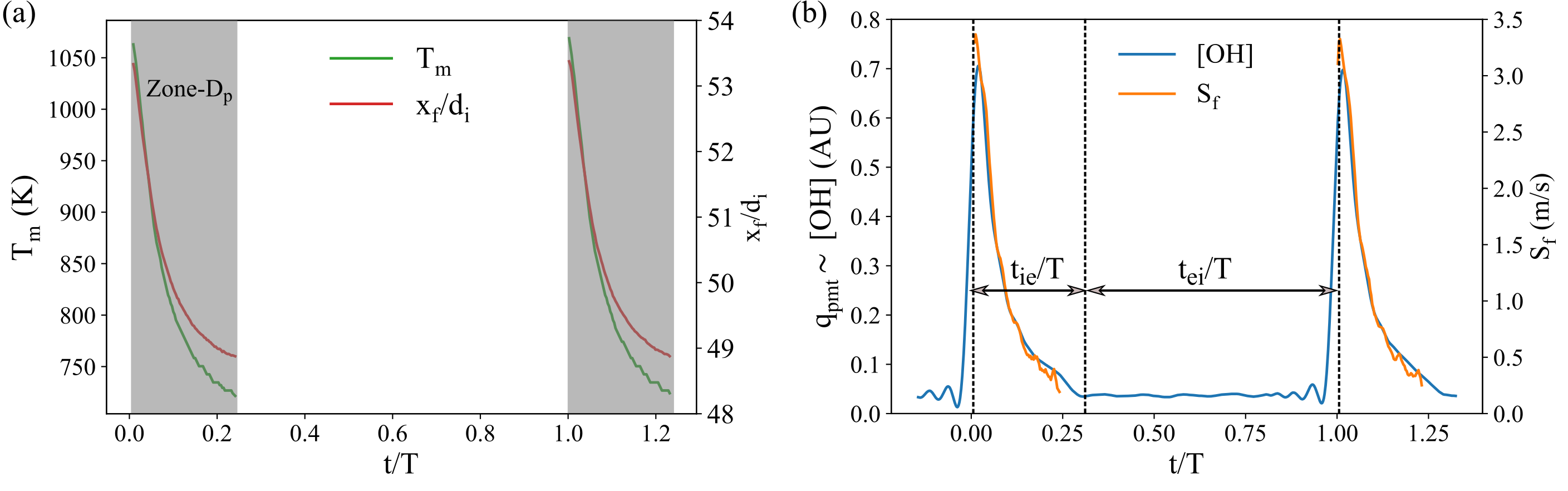}
            \caption{(a) Flame position is plotted alongside the corresponding mean flow temperature of the unburnt reactants in a typical FREI cycle. (b) OH* chemiluminescence signal is plotted alongside flame propagation speed ($S_{f}$). The plots correspond to $Re = 48$ at an equivalence ratio of $1.0$. In the figure, $t_{ie}$ represents the ignition-to-extinction timescale, $t_{ei}$ denotes the flame re-ignition timescale, and $T$ is the time period of FREI oscillations.}
            \label{fig:FREI_OH_X_SL_bs}
        \end{figure}

        Figure \ref{fig:FREI_OH_X_SL_bs} presents the typical OH* chemiluminescence signal, flame position ($x_f$), and propagation speed ($S_f$) of a FREI cycle. Ignition is marked by simultaneous peaks in OH* chemiluminescence and flame speed profiles and is followed by a decay in both signals as the flame propagates upstream. The flame eventually extinguishes after traversing a characteristic distance.
        
        The observed flame dynamics can be understood using the analytical model for flame propagation in narrow channels developed by \cite{daou_influence_2002}.

        \begin{gather*}
            V^2\ln{\left(V\right)}=\ -\kappa \tag{7}
        \end{gather*}
        
        The model non-dimensionalizes flame propagation speed as $V=\frac{2}{3}\left(\frac{1}{\bar{u}}\left|\frac{dx_f}{dt}\right|+1\right)$, which can be re-written in terms of $S_{f}$ as $V=\frac{2}{3}\left(\frac{S_f}{\bar{u}}\right)$. The heat losses at the combustor walls are accounted for using a non-dimensional parameter, $\kappa$, defined as,

        \begin{gather*}
            \kappa=\frac{Eq_w^{\prime\prime}l_f^2}{R_o^2T_a^2r_i} \tag{8}
        \end{gather*}
        
        In the above equation, $E$ is the activation energy of the one-step reaction that describes the chemical activity, $q_{w}^{\prime\prime}$ is the heat flux at the inner walls of the combustor tube, $l_{f}$ describes the flame thickness, $R_{o}$ is the universal gas constant, and $T_{a}$ is the adiabatic flame temperature. The wall heat flux, $q_{w}^{\prime\prime}$, can be expressed as $\bar{h}\left(T_a-T_{w,i}\right),$ where $\bar{h}$ is the effective heat transfer coefficient. The plots in Figure \ref{fig:Tm_all}(a) depict that the inner wall temperatures are close to the mean flow temperatures at a given axial location. We can thus approximate $\left(T_a-T_{w,i}\right)$ as $\left(T_a-T_m\right)$ in the expression for $q_{w}^{\prime\prime}$. $\left(T_a-T_m\right)$ can be further scaled as $\frac{Q_RY_F}{C_p}$ based on energy balance between the reactants and products (\cite{law_2006, daou_influence_2002}). Thus, $\kappa$ reduces to,
        
        \begin{gather*}
            \kappa=E\bar{h}\left(\frac{Q_RY_F}{C_p}\right)\frac{l_f^2}{R_o^2T_a^2r_i} \tag{9}
        \end{gather*}

        As evident from the above equation, for a fixed value of Reynolds number and equivalence ratio, $\kappa$ scales inversely with $T_a^2$. 

        \begin{gather*}
            \kappa \propto \frac{1}{T_a^2} \tag{10}
        \end{gather*}
        
        As the FREI flame auto-ignites in the primary heating zone and propagates upstream, it encounters reactants with progressively decaying mean flow temperatures (see Fig. \ref{fig:FREI_OH_X_SL_bs}(a)). Since $T_a$ scales as $\left(T_m+\ \frac{Q_RY_F}{C_p}\right)$, $T_a$ drops during the propagation phase, increasing $\kappa$. This reduces $V$ as per Equation (7), causing $S_f$ to drop during the propagation phase, and this is evident in Fig. \ref{fig:FREI_OH_X_SL_bs}(b). The plot in supplementary figure S3 depicts the inverse dependence of $V$ on $\kappa$. Given that the flame speed ($S_f$) scales with the reaction rate (\cite{law_2006}), $\omega$, a corresponding decay is expected in $\omega$, and is corroborated by the OH* chemiluminescence signal of the flame (Fig. \ref{fig:FREI_OH_X_SL_bs}(b)), which also scales with the reaction rate.
        
        Following the analytical model outlined in Equation (7) and the plot in supplementary figure S3, the non-dimensional heat loss parameter, $\kappa$, increases and reaches a local maximum of 0.18 as the upstream propagating flame nears extinction (\cite{daou_influence_2002}). This is accompanied by a simultaneous decay in $V$, which drops to a minimum of 2/3 close to extinction. At this point, $\frac{dx_f}{dt}$ will reduce to zero (as per the definition of $V$), causing the flame to cease propagation and extinguish.
        
        \begin{figure} 
            \centering
            \includegraphics [width=0.9\linewidth] {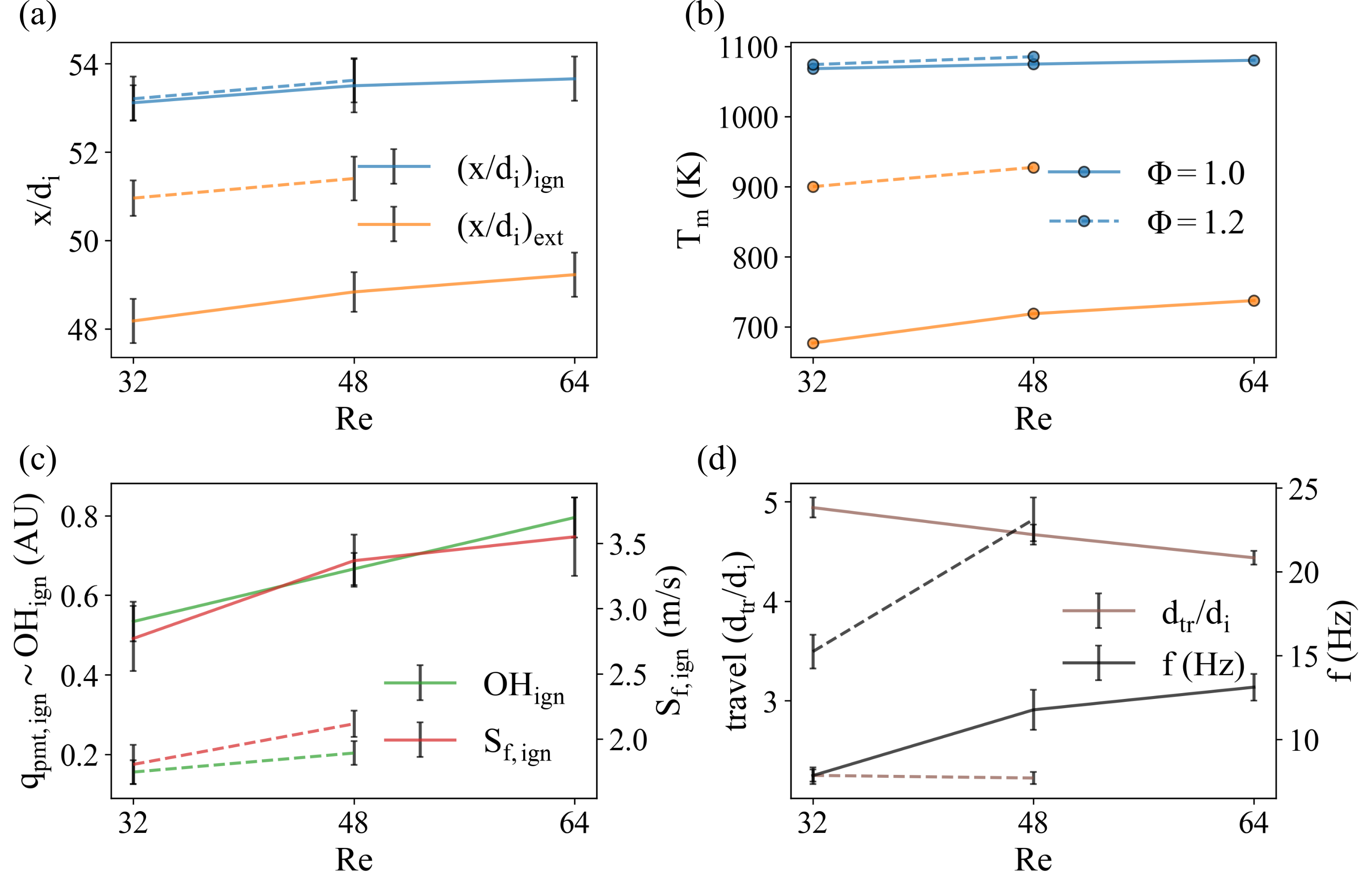}
            \caption{(a,b) Ignition and Extinction location are plotted alongside the corresponding mean flow temperature of the unburnt reactants across the space of $Re$ at which FREI is observed. (c) $(OH)_{ign}$ is plotted alongside the flame propagation speed at ignition $S_{f,ign}$. (d) Flame travel distance and repetition frequency of FREI cycles are plotted for different values of $Re$. In the figure, solid lines correspond to stoichiometric conditions, and dotted lines correspond to $\Phi=1.2$.}
            \label{fig:FREI_Ign_X_Tm_OH_Tr}
        \end{figure}
        
        Figure \ref{fig:FREI_Ign_X_Tm_OH_Tr} presents the ignition, extinction, flame speed, and OH* chemiluminescence characteristics of the baseline FREI regime as a function of the premixture Reynolds number. As $Re$ increases at a given equivalence ratio, the reactant mass flow rate increases. Mathematically, this can be expressed as,

        \begin{gather*}
            \dot{m}=\ \rho_u\left(\pi r_i^2\right)\bar{u}=\ \rho_u\left(\pi r_i^2\right)\left(Re\frac{\nu}{2r_i}\right)\ \sim \left(MW\right)_r\omega \tag{11}
        \end{gather*}
        
        The above equation implies that the reaction rate increases with an increase in the Reynolds number ($\omega\ \sim Re$) to supplement the higher consumption rate of the reactants at higher $Re$. This trend is corroborated by the plots in Fig. \ref{fig:FREI_Ign_X_Tm_OH_Tr}(c), which show that the OH* chemiluminescence signal of the flame at ignition increases with an increase in the Reynolds number. $S_{f}$ at ignition exhibits a similar trend against $Re$ owing to the direct correspondence between $\omega$ and $S_f$ (Fig \ref{fig:FREI_Ign_X_Tm_OH_Tr}(c)). It is interesting to note that the mean flow temperature of the reactants at ignition follows a similar trend, and increases with $Re$ (Fig. \ref{fig:FREI_Ign_X_Tm_OH_Tr}(b)). Given that $T_a$ scales as $\left(T_m+\ \frac{Q_RY_F}{C_p}\right)$, an associated increase in $T_{a}$ is anticipated. This increment in $T_a$ with increasing $Re$ positively contributes to the required increase in the reaction rate ($\omega$) at higher Reynolds numbers (alongside the contribution due to increased concentrations of the reactants at higher $Re$), as demonstrated in the equation below.

        \begin{gather*}
            \omega\ \sim AT_a^b\exp{\left(-\frac{E_a}{R_oT_a}\right)}\left[Fuel\right]^c\left[Air\right]^d \tag{12}
        \end{gather*}

        However, as illustrated in section \ref{Mean flow Temperature} that formulates the heat transfer from the walls into the fluid, the mean flow temperature ($T_m$) of the reactants at a given axial distance ($x$) drops with increasing $Re$ (Figure \ref{fig:Tm_all}(a)). Thus, to attain higher mean flow temperatures at higher $Re$, the reactant mixture has to travel further downstream into regions of higher wall temperatures. This trend is evident in Fig. \ref{fig:FREI_Ign_X_Tm_OH_Tr}(a), which demonstrates that the ignition location moves downstream with increasing $Re$. 
        
        Once auto-ignition is achieved, the flame propagates upstream with a decaying profile in $S_f$ and $V$ (Fig. \ref{fig:FREI_OH_X_SL_bs}(b)). Equation (7) can be used to estimate the decay rate (drop in $V$ against $x$), $\frac{dV}{dx}$, as,

        \begin{gather*}
            \frac{dV}{dx}=\frac{1}{V\left(1+2\ln{\left(V\right)}\right)}\left(\frac{2c}{T_a^3}\right)\left(\frac{dT_a}{dx}\right) \tag{13}
        \end{gather*}

        where, $c=E\bar{h}\left(\frac{Q_RY_F}{C_p}\right)\frac{l_f^2}{R_o^2r_i}$ is a constant. It should be noted that in the above equation $\left(\frac{dT_a}{dx}\right) = \left(\frac{dT_m}{dx}\right)$ following the definition of $T_a$ $\left(=\left[T_m+\ \frac{Q_RY_F}{C_p}\right]\right)$. The variation of $\frac{dV}{dx}$ with $Re$ can help us understand how the flame travel distance (distance between the ignition and extinction location) varies with the Reynolds number. To simplify the analysis, we can estimate a scale for $\frac{dV}{dx}$ by evaluating it at $x_{ign}$ (ignition location).

        \begin{gather*}
            p=\ \frac{dV}{dx} \sim \frac{c^\prime e^{g x_{ign}}}{V_{ign}(1+2\ln{\left(V_{ign}\right)})} \tag{14}
        \end{gather*}
        
        In the above equation, $c^\prime=\frac{2cg T_{m,o}}{\left(\frac{QY_F}{C_p}\right)^3}$  and $V_{ign}=V(x_{ign})$. The formulation employs an exponential profile to approximate $T_m$ as  $T_{m,o}\exp(g x)$ for $x\le x_{ign}$. This approximation is depicted in Figure. \ref{fig:Tm_all}(a). Differentiating the above equation with respect to $Re$ yields us,

        \begin{gather*}
            \frac{\partial p}{\partial(Re)}=\frac{\left[\left[V_{ign}\left(1+2\ln{\left(V_{ign}\right)}\right)c^\prime e^{g x_{ign}}\right]\left(\frac{\partial x_{ign}}{\partial\left(Re\right)}\right)-\left[c^\prime e^{g x_{ign}}\left(3+2\ln{\left(V_{ign}\right)}\right)\right]\left(\frac{\partial V_{ign}}{\partial\left(Re\right)}\right)\right]}{\left[V_{ign}(1+2\ln{\left(V_{ign}\right)})\right]^2} \tag{15}
        \end{gather*}
         
        Figure \ref{fig:FREI_OH_X_SL_bs} shows that $\left(\frac{\partial x_{ign}}{\partial\left(Re\right)}\right)>0$ and the plot in supplementary section S4 depict that $\left(\frac{\partial V_{ign}}{\partial\left(Re\right)}\right)<0$ in the FREI regime. Substituting these inequalities in the above equation yields,

        \begin{gather*}
            \frac{\partial p}{\partial(Re)}>0 \tag{16}
        \end{gather*}
        
        The above-obtained inequality can now be used to deduce the trends in the travel distance ($d_{tr}$) upon varying the Reynolds numbers. Following the definition of $p$, we can obtain the following scaling law for $d_{tr}$,

        \begin{gather*}
            p=\frac{dV}{dx} \sim \frac{V_{ign}}{d_{tr}}\Longrightarrow d_{tr} \sim \frac{V_{ign}}{p} \tag{17}
        \end{gather*}
        
        Differentiating the above scaling law with respect to Re gives,

        \begin{gather*}
            \frac{\partial\left(d_{tr}\right)}{\partial\left(Re\right)} \sim \frac{1}{p}\left(\frac{\partial V_{ign}}{\partial\left(Re\right)}\right)-\frac{d_{tr}}{p}\left(\frac{\partial p}{\partial\left(Re\right)}\right) \tag{18}
        \end{gather*}
        
        Since $\frac{\partial p}{\partial(Re)}>0$ (Equation 16) and $\left(\frac{\partial V_{ign}}{\partial\left(Re\right)}\right)<0$ (Supplementary Figure S4), the above equation yields,

        \begin{gather*}
            \frac{\partial\left(d_{tr}\right)}{\partial\left(Re\right)}<0 \tag{19}
        \end{gather*}
        
        This implies that the flame’s travel distance decreases with increasing $Re$, and the trend is evident in Fig. \ref{fig:FREI_Ign_X_Tm_OH_Tr}(d). However, since, $d_{tr}=x_{ign}-x_{ext}$, we can now estimate the expected trend of $x_{ext}$ against $Re$.

        \begin{gather*}
            \frac{\partial\left(x_{ext}\right)}{\partial\left(Re\right)}=\ \frac{\partial\left(x_{ign}\right)}{\partial\left(Re\right)}-\frac{\partial\left(d_{tr}\right)}{\partial\left(Re\right)} \tag{20}
        \end{gather*}
        
        In the above equation, $\frac{\partial\left(x_{ign}\right)}{\partial\left(Re\right)}>0$ (Figure \ref{fig:FREI_Ign_X_Tm_OH_Tr}(a)) and $\frac{\partial\left(d_{tr}\right)}{\partial\left(Re\right)}<0$ (Equation 19). This implies that

        \begin{gather*}
            \frac{\partial\left(x_{ext}\right)}{\partial\left(Re\right)}>0 \tag{21}
        \end{gather*}
        
        The inequality implies that the extinction location moves downstream as the Reynolds number increases. This is corroborated by the trends observed in Figure \ref{fig:FREI_Ign_X_Tm_OH_Tr}(a). Equation (21) also implies that the flame extinguishes at higher mean flow temperatures at higher $Re$ (since $\frac{dT_m}{dx}>0$ in the extinction region of the FREI regime) and is evident from the plots in Figure \ref{fig:FREI_Ign_X_Tm_OH_Tr}(b).
        
        The plots in Figure \ref{fig:FREI_Ign_X_Tm_OH_Tr}(d) suggest that the frequency of the FREI cycles increases with increasing Reynolds numbers. To understand this trend, let us break down the time period associated with a FREI cycle ($T$) into different time scales. A FREI cycle has three major events: Ignition, Extinction and re-ignition. Accordingly, there are two associated timescales (Fig. \ref{fig:FREI_OH_X_SL_bs}(b)): time between ignition and extinction ($t_{ie}$) and time between extinction and re-ignition ($t_{ei}$). 

        \begin{gather*}
            T=t_{ie}+t_{ei} \tag{22}
        \end{gather*}

        The timescale, $t_{ie}$, describes the time associated with the flame to propagate upstream, post-ignition, and traverse a distance of $d_{tr}$. Likewise, $t_{ei}$ describes the time associated with the fresh reactant mixture to travel a distance of $d_{tr}$ and auto-ignite at the ignition location, initiating the next FREI cycle. We can thus scale T as,

        \begin{gather*}
            \left(\frac{1}{f}\right) \sim T\ \sim \frac{d_{tr}}{\left(\frac{dx_f}{dt}\right)_{ign}}+\frac{d_{tr}}{\bar{u}} \tag{23}
        \end{gather*}

        Differentiating the above equation with respect to $Re$, we get,

        \begin{gather*}
            \frac{\partial T}{\partial\left(Re\right)}\ \sim \left(\frac{1}{\left(\frac{dx_f}{dt}\right)_{ign}}+\frac{1}{\bar{u}}\right)\left(\frac{\partial(d_{tr})}{\partial\left(Re\right)}\right)-\frac{d_{tr}}{\left(\left(\frac{dx_f}{dt}\right)_{ign}\right)^2}\left(\frac{\partial\left(\left(\frac{dx_f}{dt}\right)_{ign}\right)}{\partial\left(Re\right)}\right)-\frac{d_i d_{tr}}{{\nu Re}^2} \tag{24}
        \end{gather*}
   
        However, the plots in the supplementary figure S4 depict that $\frac{\partial\left(\left(\frac{dx_f}{dt}\right)_{ign}\right)}{\partial\left(Re\right)}>0$, while Equation (19) illustrates that $\frac{\partial(d_{tr})}{\partial\left(Re\right)}<0$. Therefore, the above equation simplifies to,

        \begin{gather*}
            \frac{\partial T}{\partial\left(Re\right)}<0 \tag{25}
        \end{gather*}
        
        The equation implies that the time period of the FREI oscillations decreases with increasing $Re$, and thus justifies the increasing trend of FREI frequencies with $Re$, depicted in Figure \ref{fig:FREI_Ign_X_Tm_OH_Tr}(d). The trends described above are variations with respect to the Reynolds number. Similar trends can be obtained against equivalence ratio. 
        
        As the equivalence ratio increases from 1.0 to 1.2, the flame thickness ($l_f$) increases and the adiabatic flame temperature ($T_a$) reduces. Thus, as per Equation (9), $\kappa$ increases, which in turn drops $V$ (Equation (7)). Thus, the flame propagation speed ($S_f$) is expected to drop with increasing $\Phi$, and is corroborated by the plots in Figure \ref{fig:FREI_Ign_X_Tm_OH_Tr}(c). The plots illustrate that both $S_f$ and OH* chemiluminescence signals at ignition drop with increasing $\Phi$. Ignition locations are comparable at the equivalence ratios of 1.0 and 1.2 across the space of Re (Figure \ref{fig:FREI_Ign_X_Tm_OH_Tr}(a)). However, the extinction location is observed to shift downstream reducing $d_{tr}$ at $\Phi=1.2$. This trend can be understood by estimating how $\frac{dV}{dx}$ changes with increasing $\Phi$. Following Equation (14), we see that,

        \begin{gather*}
            p\left(=\frac{dV}{dx}\right)\ \propto\frac{l_f^2}{\left(\frac{QY_F}{C_p}\right)^2}\left(\frac{1}{V_{ign}\left(1+2\ln{\left(V_{ign}\right)}\right)}\right) \tag{26}
        \end{gather*}
        
        As $\frac{QY_F}{C_p}$ and $V_{ign}$ tend to decrease, while $l_f$ increases as the mixture shifts from stoichiometric ($\Phi=1.0$) to fuel-rich conditions ($\Phi=1.2$), it is expected that $p$ will be higher at $\Phi=1.2$. Thus,

        \begin{gather*}
            \frac{\partial p}{\partial\Phi}>0\ \quad \text{for} \quad \Phi>1 \tag{27}
        \end{gather*}
        
        Following a similar process as depicted earlier, we can formulate $\frac{\partial\left(d_{tr}\right)}{\partial\left(\Phi\right)}$ by differentiating Equation (17) with respect to $\Phi$.

        \begin{gather*}
            \frac{\partial\left(d_{tr}\right)}{\partial\left(\Phi\right)}=\frac{1}{p}\left(\frac{\partial V_{ign}}{\partial\left(\Phi\right)}\right)-\frac{d_{tr}}{p}\left(\frac{\partial p}{\partial\left(\Phi\right)}\right) \tag{28}
        \end{gather*}
        
        However, since, $\left(\frac{\partial V_{ign}}{\partial\left(\Phi\right)}\right)<0$ (Fig. \ref{fig:FREI_Ign_X_Tm_OH_Tr}(c)) and $\left(\frac{\partial p}{\partial\left(\Phi\right)}\right)>0$ (Equation (28)) in fuel-rich conditions, the travel distance ($d_{tr}$) is expected to drop as $\Phi$ increases from 1.0 to 1.2. This causes the flame to extinguish at higher values of $x_{ext}$ (at $\Phi$=1.2), in comparison with the stoichiometric conditions. A drop in the travel distance is also accompanied by a drop in the time period of FREI oscillation, leading to an increase in the repetition frequency at $\Phi=1.2$ (Fig. \ref{fig:FREI_Ign_X_Tm_OH_Tr}(d))

    \subsubsection{Propagating Flames (PF)}\label{PF} \addvspace{10pt}

\begin{figure}
            \centering
            \includegraphics [width=0.9\linewidth] {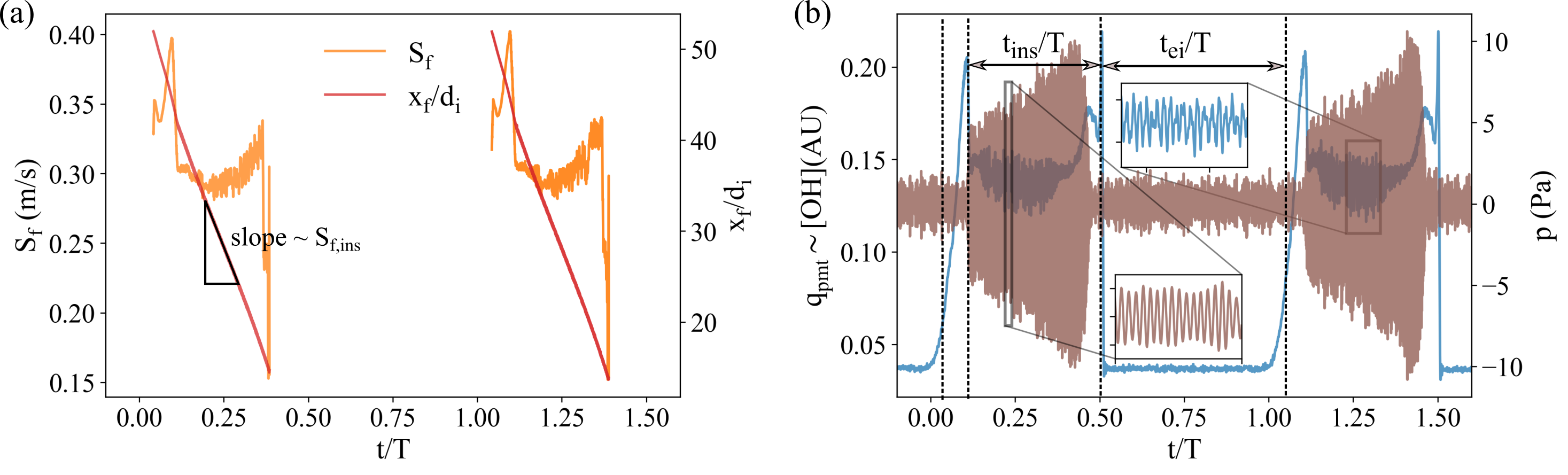}
            \caption{(a) Flame position is plotted alongside the flame propagation speed for a typical propagating flame cycle. Panel (b) plots the corresponding OH* chemiluminescence signal and the pressure signal from the microphone. The profiles correspond to the Reynolds number of $48$. In the figure, $t_{ins}$ denotes the time associated with the flame to traverse a distance of $x_{ins}$, and $t_{ei}$ is the time between extinction and re-ignition.}
            \label{fig:PF_OH_X_SL_P_bs}
        \end{figure}

\begin{figure}
            \centering
            \includegraphics [width=0.65\linewidth] {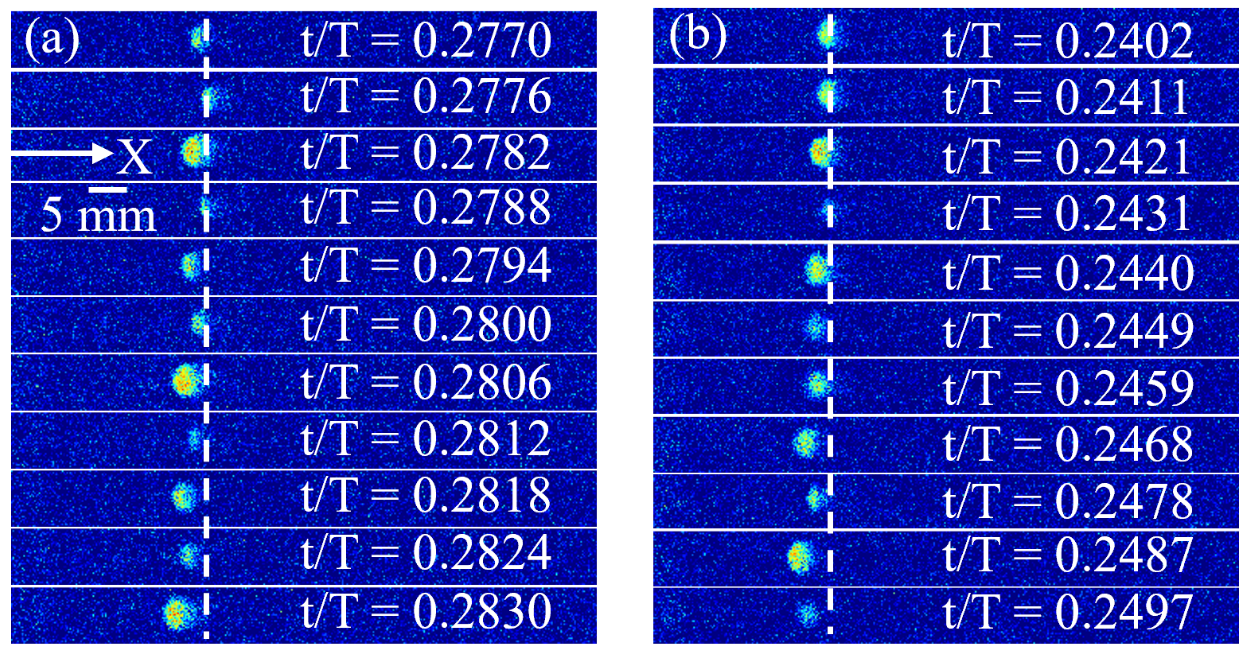}
            \caption{(a) Image sequence depicting the violent back-and-forth motion of the propagating flame at $Re=48$ for the baseline case. (b) Similar back-and-forth motion of the flame front at the same Reynolds number for $d/d_{i} = 18$}
            \label{fig:Obs_PF_Ins}
        \end{figure}

        The OH* chemiluminescence signature of a typical propagating flame, alongside its instantaneous flame location and flame propagation speed, is plotted in Fig.\ref{fig:PF_OH_X_SL_P_bs}. Similar to that observed in the FREI regime, the flame auto-ignites in the high-temperature primary heating zone and starts to propagate upstream into regions of lower mean flow temperatures. Interestingly, unlike FREI, flames in this regime develop instabilities (reflected as fluctuation in the OH* chemiluminescence signal) as they propagate upstream. These instabilities grow in magnitude, turn violent over the period (Fig.\ref{fig:PF_OH_X_SL_P_bs}(b)), and become evident in flame imaging as back and forth motion of the flame front while still continuing to propagate upstream (Fig. \ref{fig:Obs_PF_Ins}(a)).
        
\begin{figure}
            \centering
            \includegraphics [width=0.9\linewidth] {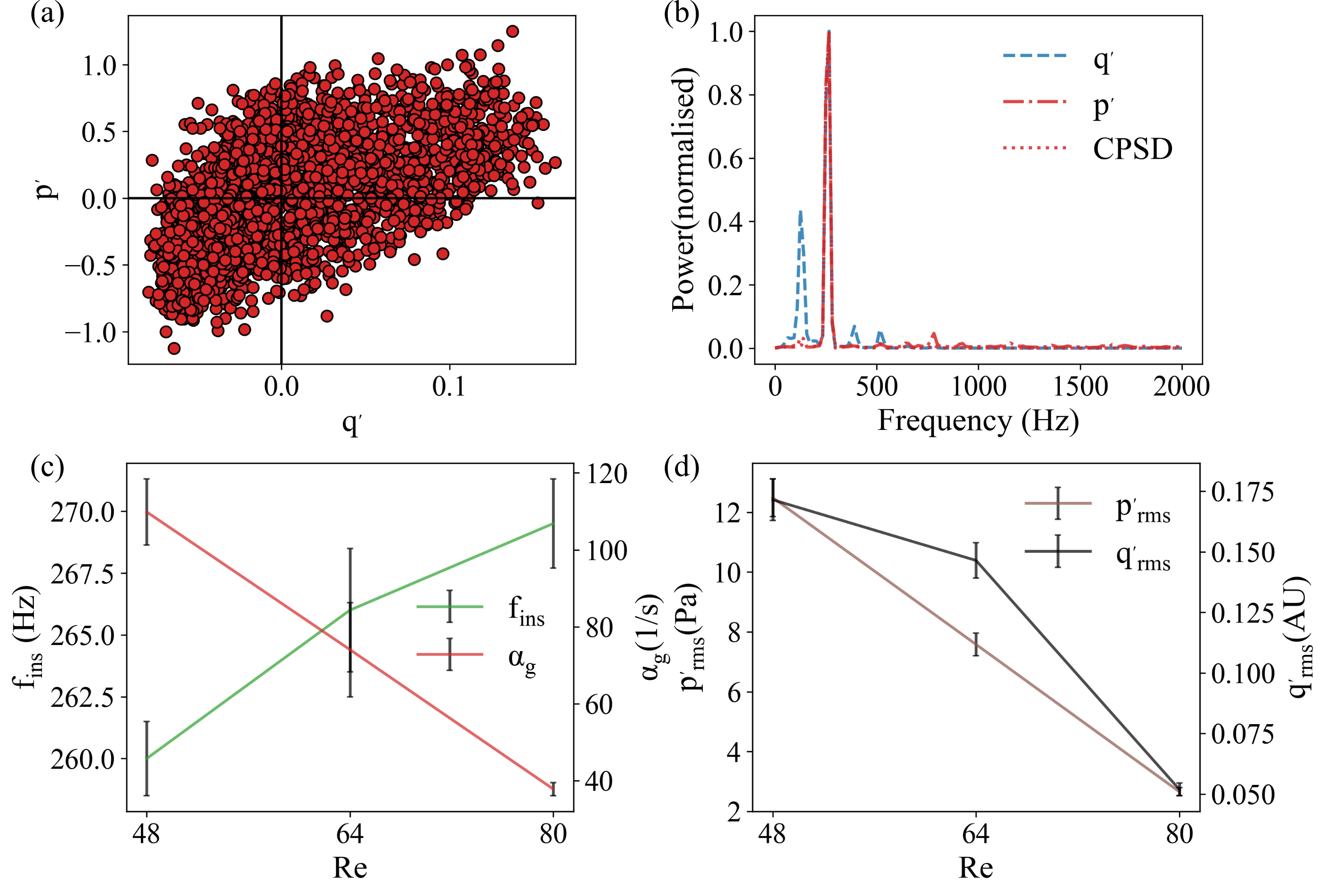}
            \caption{(a) Phase plot between $q^{\prime}$ and $p^{\prime}$ during the thermo-acoustic phase of a propagating flame. (b) FFT of the $p^{\prime}$ and $q^{\prime}$ signals is plotted alongside the cross power spectral density of $p^{\prime}$ and $q^{\prime}$ (c) Thermoacoustic frequency ($f_{ins}$) and the mean growth rate of the thermoacoustic instability is plotted against Reynolds number. (d) The root mean square value of the pressure fluctuations and the PMT fluctuations are plotted across the space of $Re$. It should be noted that the pressure fluctuations presented in the figure have been corrected to account for the spatial location of the microphone with respect to the instantaneous location of the flame.}
            \label{fig:PF_OH_P_FFT_angle_bs}
        \end{figure}

        These fluctuations are accompanied by a reduction in the mean flame propagation speeds and OH* chemiluminescence signal (Fig.\ref{fig:PF_OH_X_SL_P_bs}(b)). When decomposed in the frequency space, it becomes evident that these fluctuations in the OH* chemiluminescence signature exhibit a distinct monotone frequency close to the natural harmonic of the combustor tube (Fig. \ref{fig:PF_OH_P_FFT_angle_bs}(b)). Interestingly, the microphone captures pressure fluctuations at the same frequency (Fig.\ref{fig:PF_OH_X_SL_P_bs}(b) and Fig. \ref{fig:PF_OH_P_FFT_angle_bs}(b)). A phase plot between heat release rate (OH* chemiluminescence fluctuations) and pressure fluctuations reveals that these fluctuations are coupled (Fig. \ref{fig:PF_OH_P_FFT_angle_bs}(a,b)) and that the heat release rate fluctuations tend to lag behind the pressure fluctuations. For the plots corresponding to Fig. \ref{fig:PF_OH_P_FFT_angle_bs}(a,b), the heat release signal was found to lag behind the pressure signal by a phase angle of 30.8 degrees (estimated from the cross-power spectral density at the frequency corresponding to the peak power density, presented in Fig. \ref{fig:PF_OH_P_FFT_angle_bs}(b)). It is important to note that in the plots presented in Fig. \ref{fig:PF_OH_P_FFT_angle_bs}, $p^{\prime}$ and $q^{\prime}$ represent the fluctuating components of the pressure and PMT signals, respectively, which were obtained by subtracting the moving mean values of these signals from the original data.

        The plots in Fig. \ref{fig:PF_OH_P_FFT_angle_bs}(a,b) make it clear that the observed instability is thermo-acoustic in nature, in which the fluctuating heat release rate at the flame adds energy to the acoustic field, causing the pressure fluctuations to amplify, which, in turn, causes velocity fluctuations upstream of the flame and aggravate the heat release rate fluctuations. These fluctuations, however, dampen as the flame propagates to the upstream end of the combustor tube and encounters the meshed constriction, where it extinguishes. The process repeats again when the reactant mixture re-ignites with a characteristic time delay, generating a new propagating flame. The authors hypothesise that the developed thermo-acoustic instability is responsible for the propagating flame (observed at $\phi=0.8$) to sustain till the upstream end of the tube, while unsteady flames at $\phi=1.0$ and $1.2$ extinguish after a characteristic propagation distance. The instability causes the flame front to move back and forth, which enhances the mixing between the upstream reactant mixture and the product gases (Fig. \ref{fig:Obs_PF_Ins}). The resulting preheating of the reactants might explain why the flame is able to sustain till the upstream end of the tube, wherein the mean flow temperature of the upstream gases is close to 300K. 

        Fig. \ref{fig:PF_OH_P_FFT_angle_bs}(c) plots the variation of the observed thermo-acoustic frequency ($f_{ins}$) against $Re$. $f_{ins}$ is found to increase with increasing $Re$. The figure also plots the variation in the growth rate ($\alpha_g$) of the OH* chemiluminescence fluctuations which tends to attain a state of limit cycle, as depicted in Fig. \ref{fig:PF_OH_X_SL_P_bs}(b). $\alpha_g$ tends to decrease with increasing $Re$. However, the decay rate of the fluctuations ($\alpha_d$) was found to remain at a near-constant value close to -8.23 (with a standard deviation of 1.2) across the space of $Re$. The constancy of the decay rate suggests that meshed constriction at the upstream end of the tube, which is a geometrical parameter independent of the Reynolds number, causes the instability to decay while extinguishing the flame. It should be noted that the growth and decay rates of the OH* chemiluminescence signal were obtained by approximating the envelope of the fluctuation (obtained by Hilbert transform) using an exponential profile (= exp($\alpha t$)). It is also interesting to note that alongside the growth rates, the RMS value of the fluctuations also tends to drop with increasing $Re$ (Fig. \ref{fig:PF_OH_P_FFT_angle_bs}(d)).
        
        \begin{figure}
            \centering
            \includegraphics [width=0.9\linewidth] {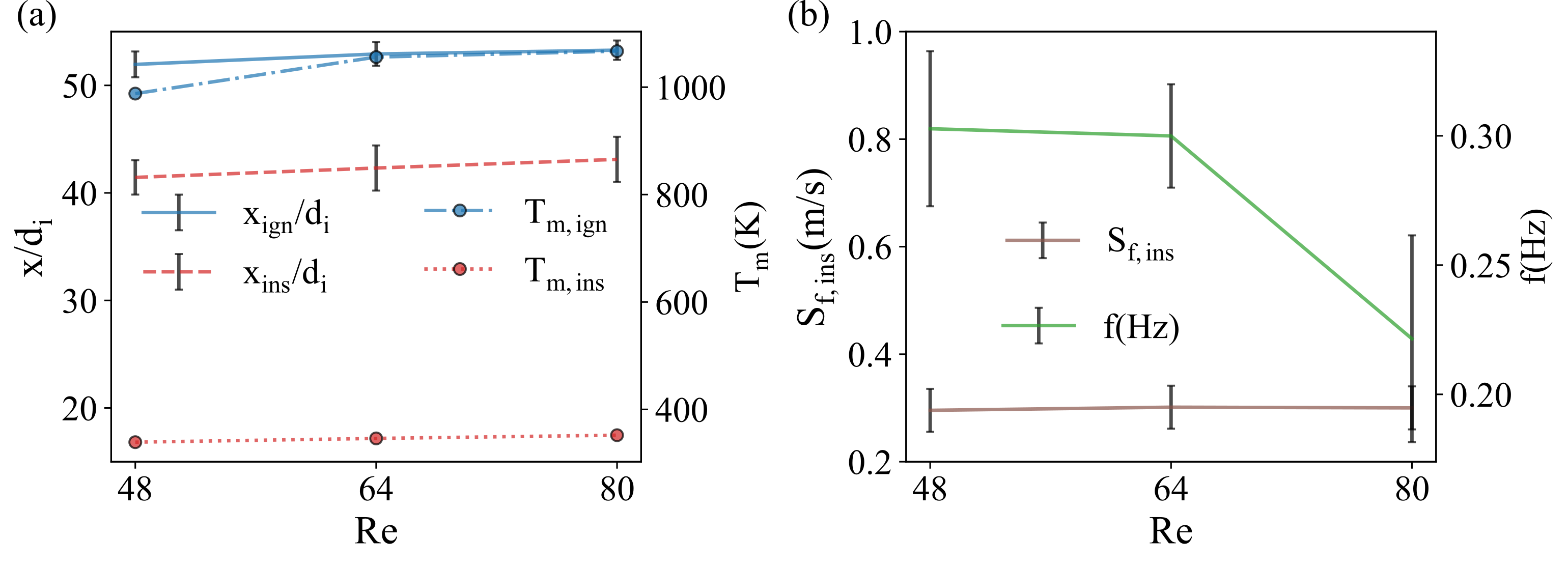}
            \caption{(a) Ignition locations and thermo-acoustic coupling locations are plotted alongside their corresponding mean flow temperatures across $Re$, for the baseline configuration. (b) The mean flame propagation speed, for $x \leq x_{ins}$, is plotted against $Re$, alongside the frequency of repetition of propagating flame cycles.}
            \label{fig:PF_Ign_Ins_SL_bs}
        \end{figure}
        
        The variation of the flame characteristics with Reynolds number is similar to those observed in the FREI regime. As the Reynolds number increases, the flame tends to ignite at higher mean flow temperatures, causing the ignition location to shift downstream (Fig. \ref{fig:PF_Ign_Ins_SL_bs}(a)). Concurrently, the spatial location where the propagating flame develops thermoacoustic instability ($x_{ins}$) also shifts downstream with increasing Reynolds numbers (Fig. \ref{fig:PF_Ign_Ins_SL_bs}(a)). The mean propagation velocity ($\bar{S}_{f,ins}$) at which the flame propagates beyond $x_{ins}$ ($x<x_{ins}$) is found to remain relatively constant with a weak increment with $Re$.
        
        A timescale analysis, analogous to that presented for the FREI regime can be used to estimate the trends in the repetition frequency (number of PF cycles per second) associated with propagating flames with increasing $Re$. The OH* chemiluminescence signature of the propagating flame presented in Fig.\ref{fig:PF_OH_X_SL_P_bs}(b), clearly reveals two dominant timescales: one associated with the propagation of the flame from $x_{ins}$ to the upstream end of the combustor tube ($t_{ins}$) and the other associated with the re-ignition of the fresh reactant mixture initiating the next PF cycle, post-extinction of the present cycle ($t_{ei}$). It is to be noted that the timescale associated with the flame to propagate from $x_{ign}$ to $x_{ins}$ is negligible in comparison with $t_{ins}$ and $t_{ei}$ (Fig.\ref{fig:PF_OH_X_SL_P_bs}(b)). Scaling the dominant timescales in terms of the associated length and velocity scales, we get,

        \begin{gather*}
            T\ \sim t_{ins}+t_{ei\ } \sim \frac{x_{ins}}{\bar{\left(\frac{dx_f}{dt}\right)}_{ins}}+\frac{x_{ign}}{\bar{u\ }} \tag{31}
        \end{gather*}
        
        In the above equation, $\bar{\left(\frac{dx_f}{dt}\right)}_{ins}=\bar{S}_{f,ins}-\bar{u}$. Differentiating the equation with respect to $Re$, we get, 

        \begin{gather*}
            \frac{dT}{d\left(Re\right)} \sim \left[ \frac{1}{\bar{\left(\frac{dx_f}{dt}\right)}_{ins}}\left(\frac{dx_{ins}}{d\left(Re\right)}\right)+\frac{1}{\bar{u}}\left(\frac{dx_{ign}}{d\left(Re\right)}\right) \right] - \left[ \frac{x_{ins}}{\left(\bar{\left(\frac{dx_f}{dt}\right)}_{ins}\right)^2} \frac{d \left( \bar{\left(\frac{dx_f}{dt}\right)}_{ins} \right)}{d(Re)} + \frac{d_i x_{ign}}{\nu Re^2}\right] \tag{32}
        \end{gather*}

        It should be noted that, $\frac{d \left( \bar{\left(\frac{dx_f}{dt}\right)}_{ins} \right)}{d(Re)}$ in the above equation can be expressed as,

        \begin{gather*}
            \frac{d \left( \bar{\left(\frac{dx_f}{dt}\right)}_{ins} \right)}{d(Re)} = \frac{d \left( \bar{S}_{f,ins} \right)}{d(Re)} - \frac{\nu}{d_i} \tag{33}
        \end{gather*}
        
        It is evident from Fig. \ref{fig:PF_Ign_Ins_SL_bs}(b) that $\frac{d \left( \bar{S}_{f,ins} \right)}{d(Re)} \sim 0$. Thus, $\frac{d \left( \bar{\left(\frac{dx_f}{dt}\right)}_{ins} \right)}{d(Re)} \sim 0$ since the second term in the above equation is negligible. Additionally, the term, $\frac{d_i x_{ign}}{\nu Re^2}$, in Equation (32) is negligible in comparison with the first two terms. Thus, the expression for $\frac{dT}{d\left(Re\right)}$ reduces to, 

        \begin{gather*}
            \frac{dT}{d\left(Re\right)} \sim \frac{1}{\bar{\left(\frac{dx_f}{dt}\right)}_{ins}}\left(\frac{dx_{ins}}{d\left(Re\right)}\right)+\frac{1}{\bar{u}}\left(\frac{dx_{ign}}{d\left(Re\right)}\right) \tag{34}
        \end{gather*}
        
        Since both $\left(\frac{dx_{ins}}{d\left(Re\right)}\right)$ and $\left(\frac{dx_{ign}}{d\left(Re\right)}\right)$ are greater than zero (Fig. \ref{fig:PF_Ign_Ins_SL_bs}(a)), $T$ tends to increase with $Re$, decreasing the repetition frequency of the propagating flame cycle (Fig. \ref{fig:PF_Ign_Ins_SL_bs}(b)).

    \subsubsection{Combined Flames (CF)}\label{CF} \addvspace{10pt}

\begin{figure}
            \centering
            \includegraphics [width=0.9\linewidth] {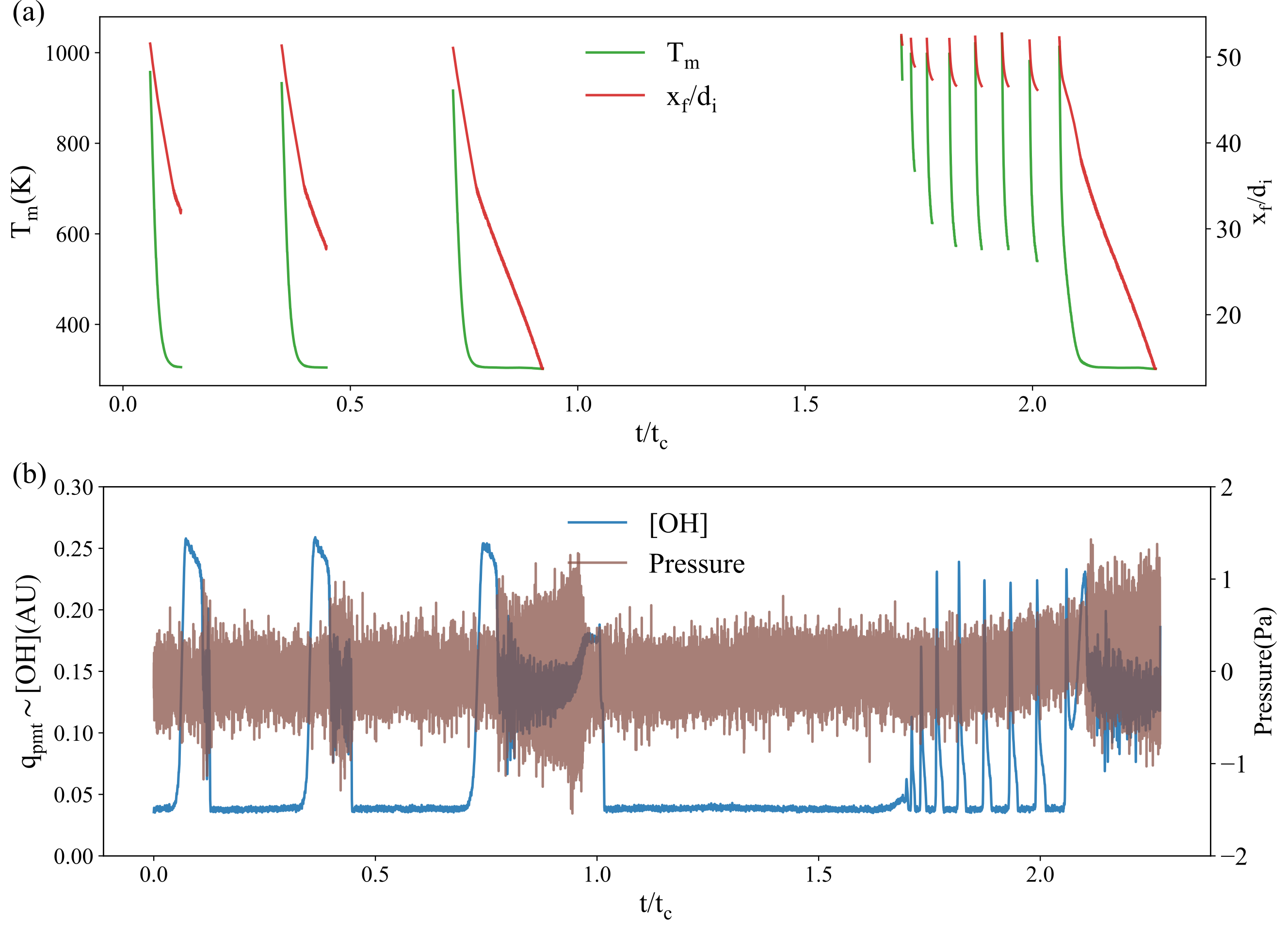}
            \caption{(a) The position of the flame is plotted alongside the corresponding mean flow temperature of the unburnt reactants, for a combined flame. Panel (b) plots the corresponding pressure signal alongside its OH* chemiluminescence signature.}
            \label{fig:CF_OH_X}
        \end{figure}

        At the Reynolds number of $32$ and equivalence ratio of $0.8$, a flame exhibiting characteristics of both FREI and PF was observed. The OH* chemiluminescence signature of the flame is plotted in Fig.\ref{fig:CF_OH_X}, alongside its corresponding instantaneous flame position and flame propagation speeds. The image sequence depicting the flame dynamics is presented in Fig.\ref{fig:Obs_All_Flames}(e).
        
        It is evident from the figures that there is no periodicity in the observed flame dynamics. Convective time scale ($t_c$) is hence used to non-dimensionalize time and represent time-series variation in Fig.\ref{fig:CF_OH_X}. The flame tends to exhibit a series of finite travel FREI cycles followed by a propagating flame cycle (wherein the flame travels to the upstream end of the combustor tube). The number of FREI cycles between consecutive propagation flame cycles was stochastic and was found not to exhibit a characteristic repetition frequency. Not only was the flame behaviour qualitatively switching between FREI and PF, but FREI descriptors (extinction locations, flame travel distance) showed quantitative changes between consecutive FREI cycles.
        
        It is to be noted that the flame regime was consistently observed at $Re=32$ and $\Phi=0.8$, across three independent experimental trials, even after extended waiting periods in each trial. As the observation of this flame type was limited to a single data point ($Re=32$, $\Phi=0.8$), further data analysis was not possible to establish any trends with respect to $Re$. This requires an independent study that deals with mesoscale flame dynamics at $Re<32$.

    \subsection{Effect of the secondary heater}\label{sec_heater}
    This section details the changes induced in the dynamics of the unsteady flame regimes due to the introduction of the secondary heater.
    
    \subsubsection{Flames with Repetitive Extinction and Ignition}\label{D_FREI} \addvspace{10pt}

\begin{figure}
            \centering
            \includegraphics [width=0.9\linewidth] {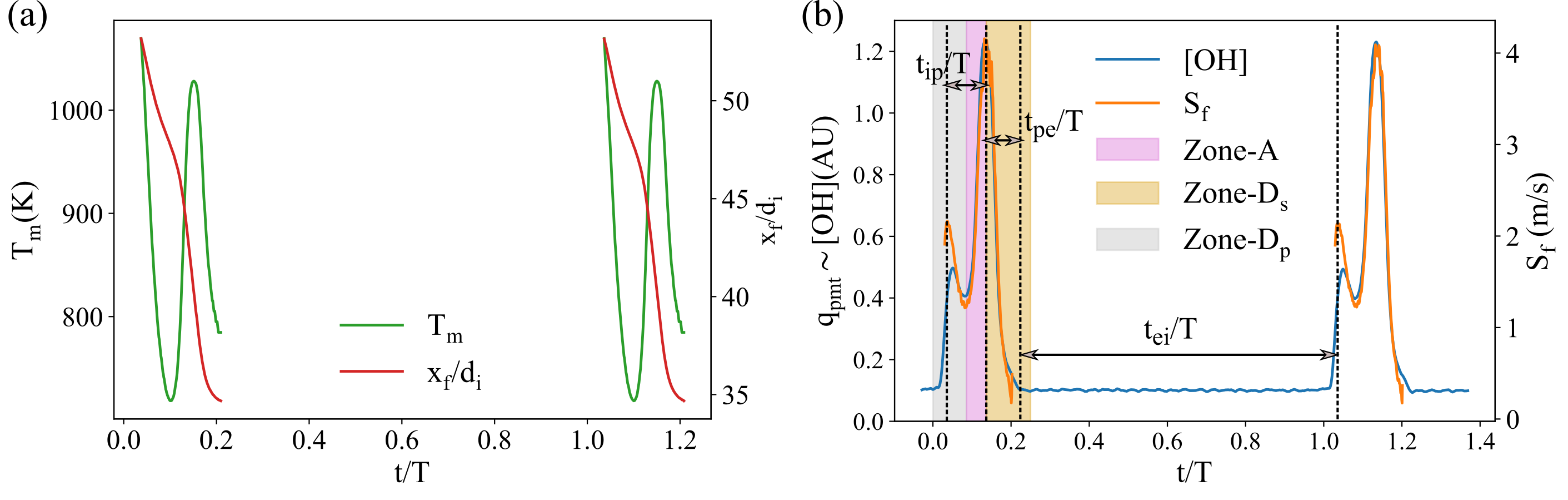}
            \caption{(a) The position of the flame is plotted alongside the corresponding mean flow temperature of the unburnt reactants, for a typical diverging FREI. (b) The flame propagation speed is plotted alongside its OH* chemiluminescence signature. The plots correspond to $Re=48$ and equivalence ratio of $1.0$, for $d/d_{i} = 15$.}
            \label{fig:FREI_OH_X_SL_bifur}
        \end{figure}

        The introduction of the flat flame (secondary heater) divides the FREI regimes into two sub-regimes: one that retains the qualitative features of the baseline FREI and another termed the Diverging-FREI (D-FREI) regime. The D-FREI regime is characterised by distinct OH* chemiluminescence and flame speed profiles which include a post-ignition peak deviating from the baseline behaviour (Fig. \ref{fig:FREI_OH_X_SL_bifur}(b)). D-FREI was observed in a characteristic range of Reynolds number ($Re$), equivalence ratio ($\Phi$), and separation distances ($d$). To understand the presence of D-FREI and the reasons for the deviation in the OH* chemiluminescence signal from its baseline behaviour, we will first examine the ignition-extinction characteristics of the flame.

\begin{figure}
            \centering
            \includegraphics [width=0.9\linewidth] {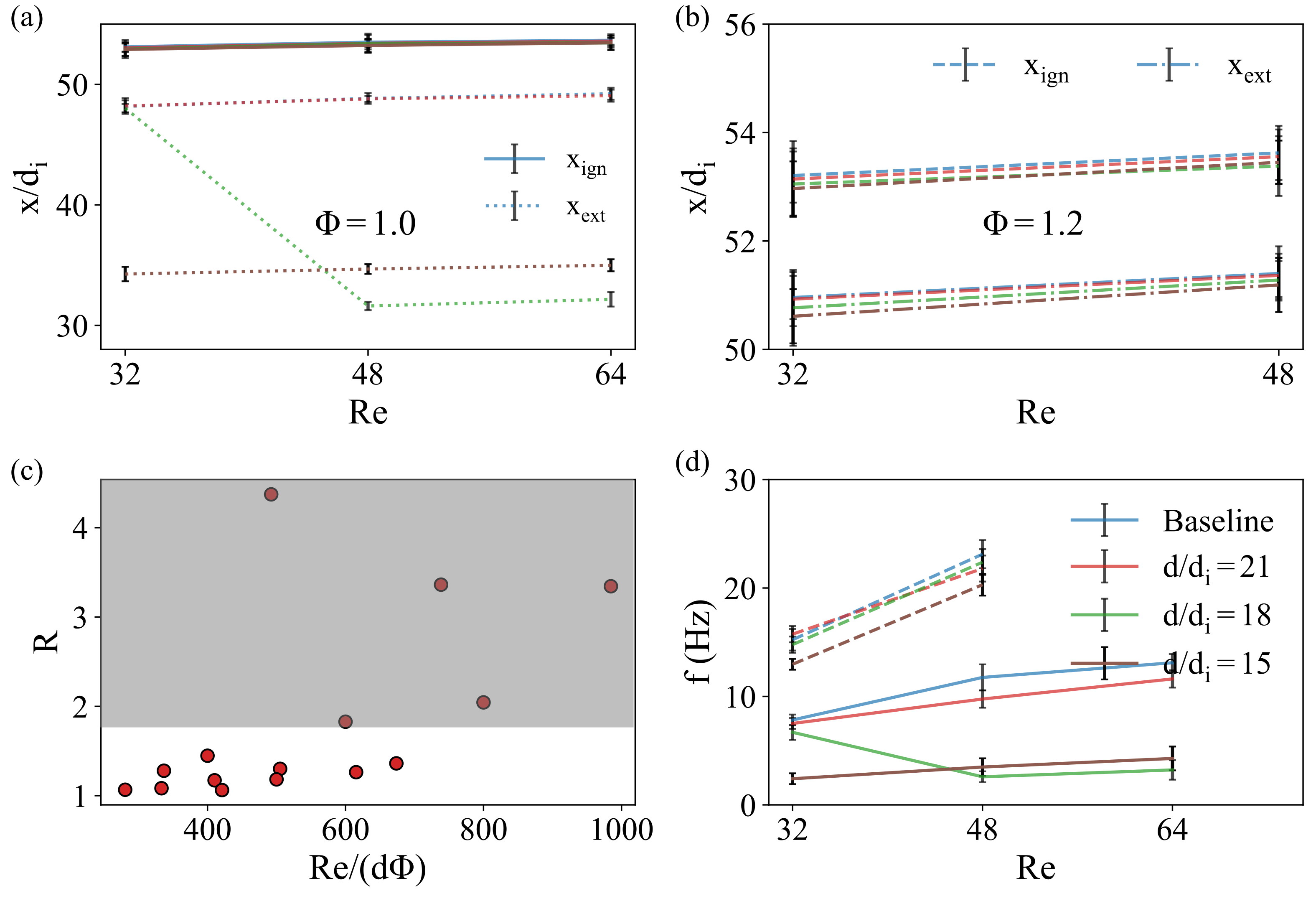}
            \caption{(a)Ignition and extinction locations are plotted against Reynolds numbers for different wall heating conditions at the equivalence ratio of $1.0$. (b) $x_{ign}$ and $x_{ext}$ are traced out for different values of $Re$ and $d/d_{i}$ at the equivalence ratio of $1.2$. (c) The parameter $R$ is plotted against $Re/(d\Phi)$ across the FREI regime for all values of $d/d_{i}$. (d) The FREI repetition frequency is plotted against $Re$ for all values of $d/d_{i}$. The solid lines in this plot correspond to $\Phi=1.0$, while the dashed lines correspond to $\Phi=1.2$.}
            \label{fig:FREI_Ign_Ext_bifur}
        \end{figure}

        As the secondary heater was introduced at different separation distances ($d$), ignition locations (Fig.\ref{fig:FREI_Ign_Ext_bifur}, plotted in solid lines) did not appear to change significantly, as ignition remained confined to the primary heating zone, which was unaffected by the addition of the secondary heater. However, the variations in the extinction locations (Fig.\ref{fig:FREI_Ign_Ext_bifur}(a), plotted in dotted lines) were prominent when the flame deviated from its baseline behaviour to exhibit D-FREI. It is to be noted that D-FREI  was observed only in a characteristic range of $Re$ and $d$ at stoichiometric conditions (Fig.\ref{fig:FREI_Ign_Ext_bifur}(a,b)). At $d/d_i=18$, the regime was observed for $Re\geq48$, and for $d/d_i=15$, D-FREI was observed across the space of $Re$. In the rest of the parametric space, the characteristics were comparable to the baseline FREI. 

        The plots in Fig.\ref{fig:FREI_Ign_Ext_bifur}(a) depict that, in the D-FREI regime, the flame travels beyond its baseline extinction location (extinction location at the corresponding $Re$ and $\Phi$ in the baseline configuration) and extinguishes further upstream beyond the secondary heating zone. An image sequence depicting the corresponding flame dynamics is presented in Fig.\ref{fig:Obs_All_Flames}(c). To comprehend this, let's assess how the introduction of the secondary heater at different separation distances changes the mean flow temperatures and the corresponding reaction rates at the baseline extinction location ($x_{ext,b}$). For this, let us define a parameter, $R$ as,

        \begin{gather*}
            R=\left[\frac{\exp{\left(-\frac{E_a}{R_oT_{m,d}\left(x_{ext,b}\right)}\right)}}{\exp{\left(-\frac{E_a}{R_oT_{m,b}\left(x_{ext,b}\right)}\right)}}\right]_{Re,\Phi}\ \ \tag{35}
        \end{gather*}
        
        In the above equation, $T_{m,d}(x_{ext,b})$ is the mean flow temperature at the baseline extinction location when the secondary heater is at a separation distance of $d$, and $T_{m,b}(x_{ext,b})$ is the mean flow temperature at the extinction location in the baseline configuration, both evaluated at a fixed value of $Re$ and $\Phi$. The variation of $R$ across the space of $Re$, $\Phi$ and $d$ is plotted in Fig.\ref{fig:FREI_Ign_Ext_bifur}(c). The parameter $R$ serves as an indicator of the extent to which the ability of a mixture to sustain combustion at $x_{ext,b}$ has increased due to the introduction of the secondary heater at a separation distance of $d$.
        
        The figure clearly shows that for cases where the flame behaviour remains consistent with the baseline behaviour, $R$ stays below approximately 1.4. In contrast, cases exhibiting diverging behaviour (D-FREI) have $R$ values ranging from 1.9 to 4.2. The map distinctly divides the space into two regions, indicating that if the value of $R$ exceeds a certain cutoff, the flame is likely to sustain combustion reactions and diverge from baseline behaviour to exhibit D-FREI. A similar map can be created if $R$ is evaluated based on the adiabatic flame temperature corresponding to $T_{m,d}(x_{ext,b})$ and $T_{m,b}(x_{ext,b})$.
        
        When the flame exhibits D-FREI behaviour, it propagates beyond the baseline extinction location into the secondary heating zone. Here, the flame traverses two distinct regions: Region $A$, characterized by progressively increasing mean flow temperatures of the upstream reactant mixture (Fig. \ref{fig:FREI_OH_X_SL_bifur}(a)), followed by Region $D_s$, wherein the mean flow temperatures progressively decrease (Fig. \ref{fig:FREI_OH_X_SL_bifur}(a)). These are additional zones introduced by the secondary heater. In Region $A$, since the upstream mixture temperature increases along the propagation direction, we expect a proportional trend in the flame temperature, flame speeds, and reaction rates, and this is reflected as a coupled spike in the OH* chemiluminescence and the flame speed plots in Fig.\ref{fig:FREI_OH_X_SL_bifur}(b). The flame attains its absolute peak in the OH* chemiluminescence and the flame speed profiles close to the exit of Region $A$ (Fig.\ref{fig:FREI_OH_X_SL_bifur}(b)). This peak is greater in magnitude than the one attained post-ignition. As the flame continues propagating into region $D_{s}$, the OH* chemiluminescence signal and the flame propagation speeds start to decay (Fig.\ref{fig:FREI_OH_X_SL_bifur}(b)) as the flame encounters reactants with progressively decaying mean flow temperatures similar to that observed in the propagation phase of the baseline case, and finally extinguishes after a characteristic travel distance. 
        
        Extending these observations of the ignition-extinction characteristics, D-FREI is bound to have a change in its characteristic time scales in comparison with its baseline counterpart. It is evident from the OH* chemiluminescence profile (Fig.\ref{fig:FREI_OH_X_SL_bifur}(a)) that flame divergence from its baseline behaviour introduces an additional time scale, $t_{ip}$, which characterises the time between ignition and the moment the flame attains its global peak in OH* chemiluminescence and $S_f$ profiles. However, a comparison of the different time scales involved reveals that the time between extinction and re-ignition ($t_{ei}$) is still the most dominant (Fig.\ref{fig:FREI_Ign_Ext_bifur}(d)), and thus the frequency variation with respect to Reynolds number is expected to follow the same increasing trend as that observed in the previous section (baseline case). However, $t_{ip}$ introduces a shift in the time period of diverging FREI, and this shift is reflected in the frequency plot in the form of a negative offset (Fig.\ref{fig:FREI_Ign_Ext_bifur}(d)).

    \subsubsection{Propagating flames}\label{PF_secondary_heater} \addvspace{10pt}

\begin{figure}
            \centering
            \includegraphics [width=0.8\linewidth] {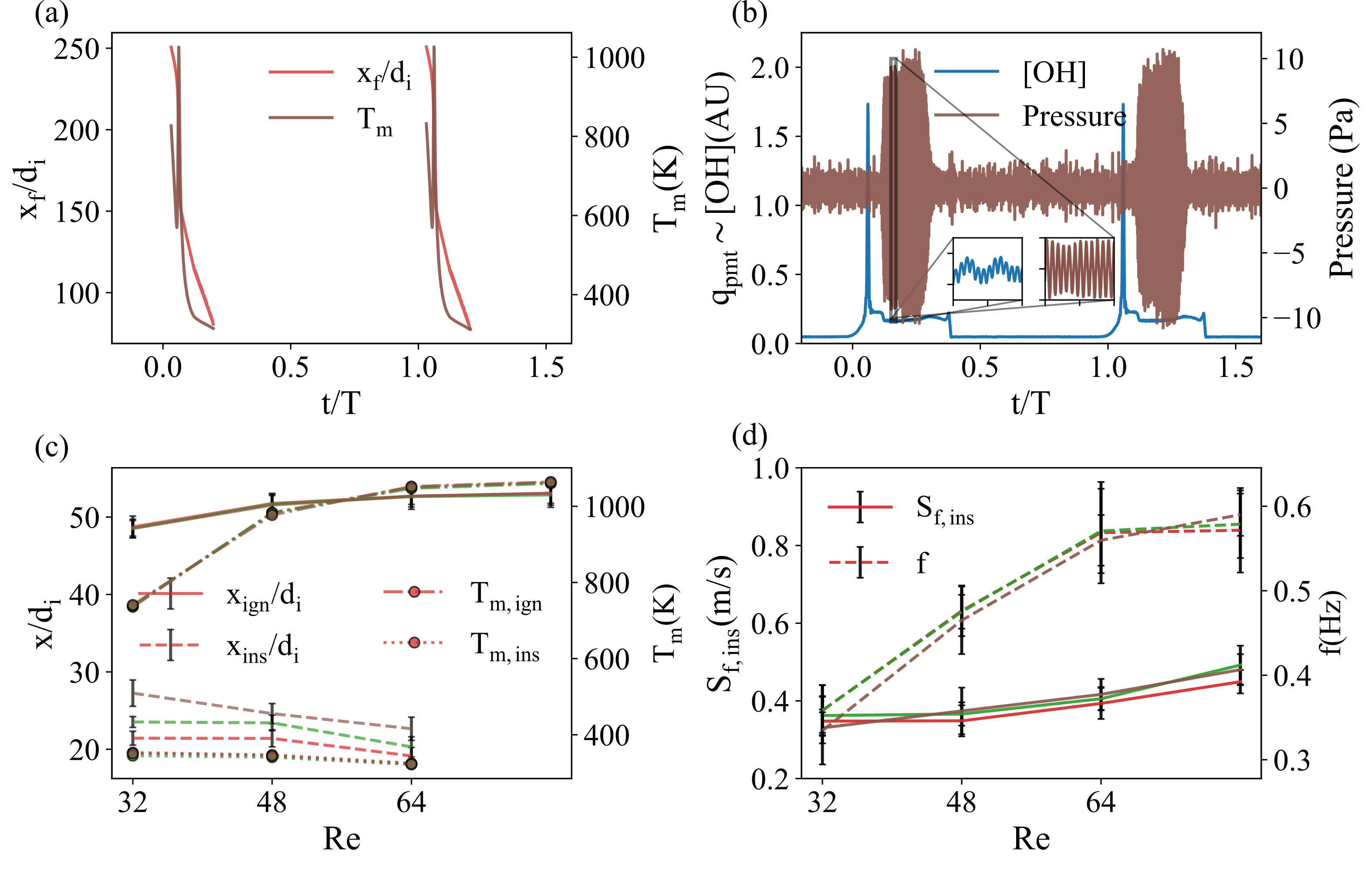}
            \caption{(a) Flame position is plotted alongside the corresponding mean flow temperature of the unburnt reactants. Panel (b) plots the corresponding OH* chemiluminescence signal and the pressure signal from the microphone. The profiles correspond to the Reynolds number of $48$ and $d/d_{i}=18$. (c) Ignition locations and thermo-acoustic coupling locations are plotted alongside their corresponding mean flow temperatures across $Re$, for all values of $d/d_{i}$. (d) The mean flame propagation speed, for $x \leq x_{ins}$, is plotted against $Re$, alongside the frequency of repetition of propagating flame cycles, across the parametric space of $d/d_{i}$.}
            \label{fig:PF_OH_X_SL_P_d}
        \end{figure}

        Propagating flames, unlike FREI, extinguish only at the upstream end of the combustor tube. Hence, they pass through the regions of $A$ and $D_{s}$ and thus exhibit a flame acceleration phase post-ignition. This is reflected in the OH* chemiluminescence and flame speed signals as a heightened peak (Fig. \ref{fig:PF_OH_X_SL_P_d}(a)).         
        
        Similar to our observations in the previous section, the ignition locations are not altered by the introduction of the secondary heater (Fig. \ref{fig:PF_OH_X_SL_P_d}(c)). However, $x_{ins}$ show significant deviation from the baseline observations. Thermoacoustic instabilities are found to affect flame propagation only after the flame crosses the secondary heating zone, implying that $x_{ins}$ shifts downstream due to the introduction of the secondary heater. 
        
        Alongside this, $x_{ins}$ is found to move upstream with increasing $Re$, which contrasts with the observations in the baseline case (Fig. \ref{fig:PF_OH_X_SL_P_d}(c)). This reversal in the trend in $x_{ins}$ against $Re$ might explain the reversal in the trend of PF repetition frequency (in comparison with the baseline observation), following equation (34). The repetition frequency of the PF cycle (f) is found to increase with increasing $Re$ (Fig. \ref{fig:PF_OH_X_SL_P_d}(d)). $S_{f,ins}$, however, follows a similar trend as that observed in the baseline case and tends to increase with increasing $Re$ (Fig. \ref{fig:PF_OH_X_SL_P_d}(d)).

\begin{figure}
            \centering
            \includegraphics [width=0.75\linewidth] {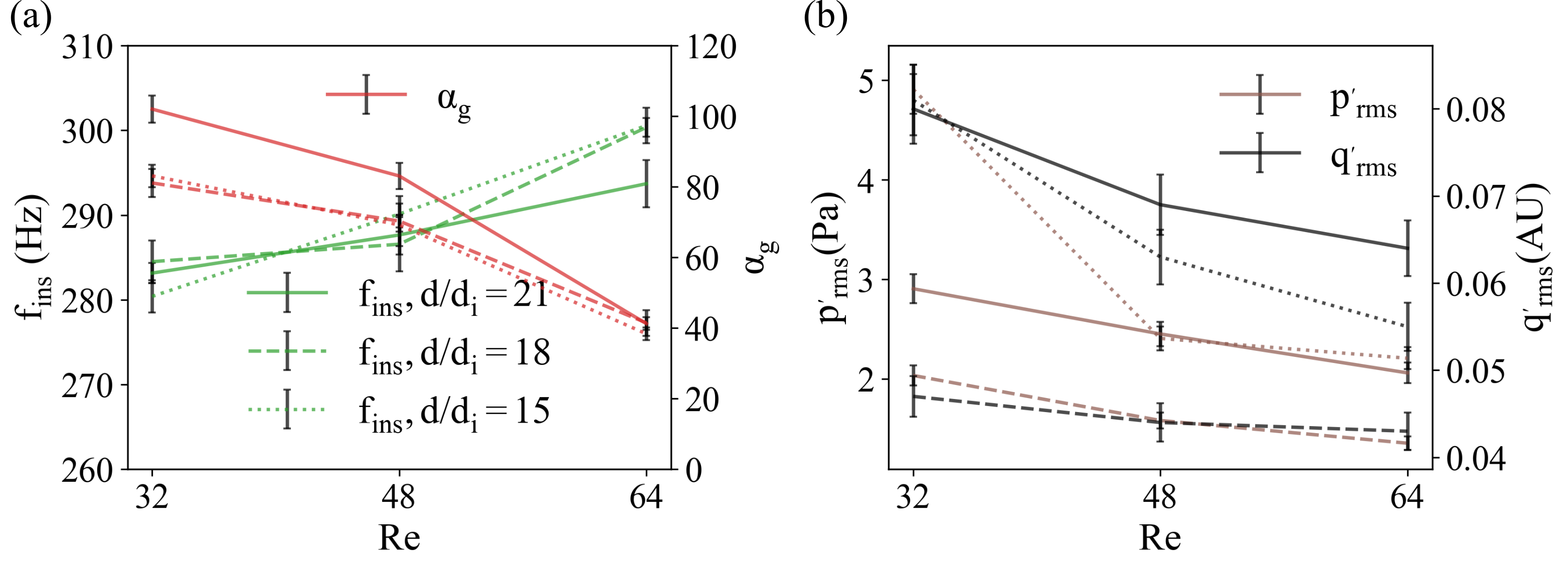}
            \caption{(a) Thermoacoustic frequency ($f_{ins}$) and the mean growth rate of the thermoacoustic instability is plotted against $Re$. (d) The root mean square value of the pressure fluctuations and the PMT fluctuations are plotted across the space of $Re$. The plots correspond to all the values of $d/d_{i}$ explored in the current work.}
            \label{fig:PF_Ign_Ins_SL_d}
        \end{figure}

        The thermoacoustic coupling frequency ($f_{ins}$) is found to increase with increasing Reynolds number (Fig. \ref{fig:PF_Ign_Ins_SL_d}(a)), while the growth rates ($\alpha_g$) and the rms value of fluctuations exhibited a decreasing trend (Fig. \ref{fig:PF_Ign_Ins_SL_d}(b)), which is synonymous with the observation in the baseline configuration. It should be noted that the fluctuations in pressure and heat release rate drop to the order of the noise level as the Reynolds number increases to $80$, and hence the corresponding values of $f_{ins}$, $p^{\prime}_{rms}$ and $q^{\prime}_{rms}$ are not reported.

        \begin{figure}
            \centering
            \includegraphics [width=0.75\linewidth] {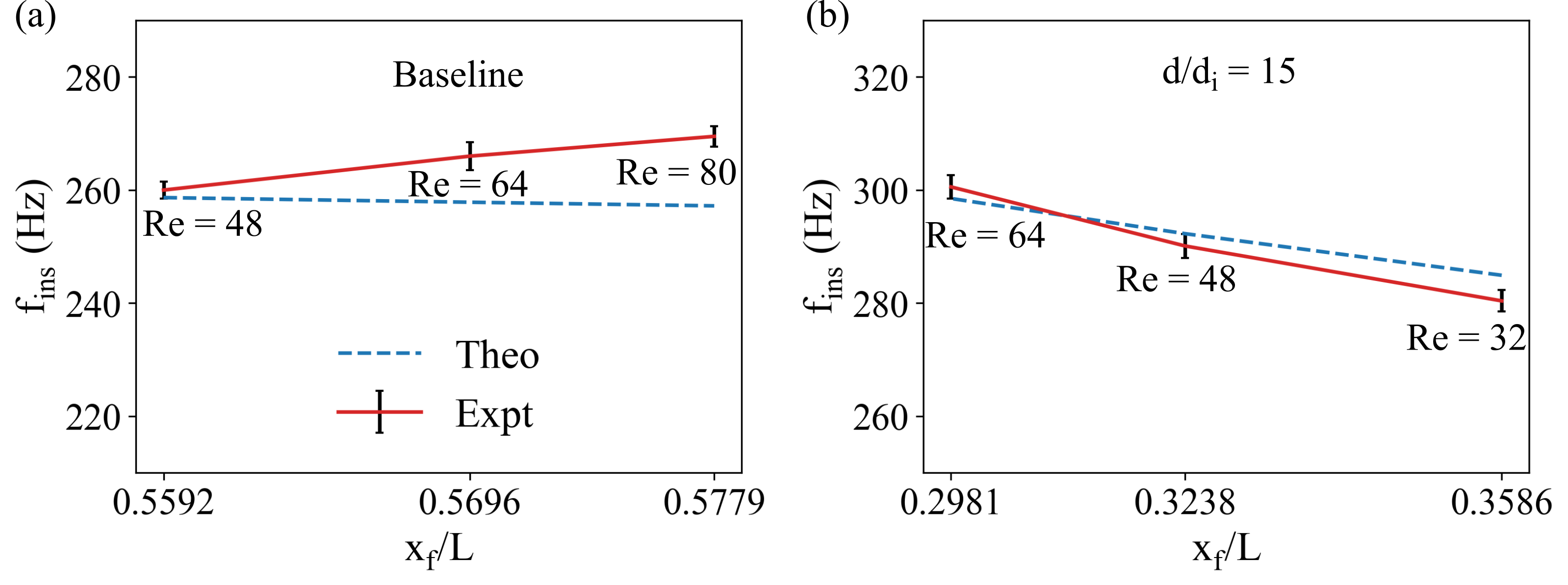}
            \caption{Theoretical estimate of the thermoacoustic frequency in comparison with the experimentally observed coupling frequency between heat release rate and pressure fluctuations. Panel (a) corresponds to the baseline case, and (b) corresponds to $d/d_{i} = 15$, respectively}
            \label{fig:PF_linear_acous}
        \end{figure}
        
        An acoustic analysis employing a passive flame model (\cite{Schuller_Poinsot_Candel_2020, MOHAN2020309}) can be used to validate our observations of $f_{ins}$ against $x_{ins}$ (or $Re$). The model assumes a temperature discontinuity across the flame front while maintaining the continuity in pressure and velocity fluctuations across it. The model was solved assuming an acoustically closed end of the inlet, and an open end at the outlet of the tube. Solving the acoustic equations under these considerations yields the following dispersion equation (\cite{Schuller_Poinsot_Candel_2020}) to evaluate $f_{ins}$ theoretically.

        \begin{gather*}
            cos\left( \frac{\omega (L-x_f)}{c_b} \right) cos\left( \frac{\omega x_f}{c_u} \right) - \sqrt{\frac{T_b}{T_u}} sin\left( \frac{\omega (L-x_f)}{c_b} \right) sin\left( \frac{\omega x_f}{c_u} \right) = 0 \tag{36}
        \end{gather*}

        In the above equation, $c_{u}$ and $c_{b}$ are the speed of sound on the unburnt and burnt side of the flame front, while $T_{u}$ and $T_b$ denote the mean temperature of the unburnt and burnt gas mixtures. Additionally, the real part of $\omega$ can be written down as $\omega_{r} = 2 \pi f_{ins}$. Figure \ref{fig:PF_linear_acous}(a,b) compares theoretical predictions of $f_{ins}$ (Equation (36)) with the experimental observations in the baseline configuration and for $d/di=15$, respectively. The predictions closely align with the experimental data.

\section{Conclusion}\label{sec:conclusion}

The study explores the dynamics of premixed methane-air flames in mesoscale channels that are subject to bimodal wall heating profiles. The separation distance between the heating peaks was varied (by varying the relative distance between two external heating sources that impose the bimodal profile), and the dynamics were investigated over a wide range of Reynolds numbers and equivalence ratios. The observed dynamics were compared against a baseline configuration that employs a single heater to impose a unimodal wall heating profile. 
Three major flame regimes were identified: a steady stationary flame regime and two unsteady flame regimes. The work is centred around these unsteady flames. Depending on the operating equivalence ratio, different types of unsteady flames were observed: one, wherein the flame goes through a succession of ignition, propagation, extinction, and re-ignition events (Flames with repetitive extinction and ignition, FREI; observed at the equivalence of $1.0$ and $1.2$), and the other, wherein the flame travels the entire length of the combustor tube (Propagating flames; observed at the equivalence ratio of $0.8$). FREI was identified between the Reynolds numbers of $32$ and $64$ at the equivalence ratio of $1.0$ and between $32$ and $48$ at the equivalence ratio of $1.2$ (at all wall temperature conditions); while propagating flames were observed for $Re \in [32, 80]$ at all bimodal heating conditions, and between $48$ and $80$ in the baseline configuration. 
The above-mentioned unsteady flame regimes repeat after a characteristic period, and the repetition frequency increases with the Reynolds number and equivalence ratio. Further trends in the ignition-extinction characteristics, OH* chemiluminescence, and flame speed profiles have been discussed in detail using a theoretical model for flame propagation in narrow channels.
FREI displays a flame bifurcation behaviour in a characteristic range of Reynolds numbers and separation distances (axial distance between the bimodal temperature peaks) at stoichiometric conditions. This divergence in the flame behaviour introduces additional peaks in their OH* chemiluminescence and flame speed profiles, altering the dynamics of the regime.
Similarly, propagating flames also tend to exhibit heightened peaks in their OH* chemiluminescence and flame speed signals when imposed with a bimodal wall heating profile (in comparison with the baseline configuration). Additionally, these flames display a characteristic thermo-acoustic coupling as they propagate upstream at frequencies close to the natural frequency of the combustor tube. The work also compares the observed thermoacoustic coupling frequency with the estimations from a linear acoustic model employing a passive flame.
The insights gained from this study can contribute to the development of optimized micro/meso scale combustors for practical applications often characterized by non-uniform wall temperatures/heat fluxes.

\section*{Declaration of competing interest} \addvspace{10pt}

The authors have no competing interest to disclose

\section*{Acknowledgments} \addvspace{10pt}

The authors would like to express their gratitude for the insightful discussions with Dr. Balasundaram Mohan from the Indian Institute of Science during this study. We also thank SERB (Science and Engineering Research Board) - CRG: CRG/2020/000055 for their financial support. S.B. acknowledges funding from the Pratt and Whitney Chair Professorship, and A.A. acknowledges the funding received through the Prime Minister's Research Fellowship scheme.

\bibliographystyle{jfm}
\bibliography{jfm-instructions}

\end{document}